\documentclass[%
preprint,
nofootinbib,
amsmath,amssymb,
aps,
tightenlines]{revtex4-1}

\usepackage{color}
\usepackage{graphicx}
\usepackage{dcolumn}
\usepackage{bm}

\usepackage{mathrsfs}

\newcommand{\n}{{\boldsymbol{n}}}

\newcommand{\U}{\underline}

\newcommand{\f}{\frac}

\newcommand{\N}{\nonumber}
\newcommand{\BF}{\begin{figure}\begin{center}}
\newcommand{\EF}{\end{center}\end{figure}}
\newcommand{\BE}{\begin{equation}}
\newcommand{\EE}{\end{equation}}
\newcommand{\BEA}{\begin{eqnarray}}
\newcommand{\EEA}{\end{eqnarray}}

\newcommand{\tr}{\textrm}
\newcommand{\IG}{\includegraphics}
\def\v#1{\boldsymbol #1}

\newcommand{\ms}{M_{\odot}}

\begin{document}

\def\a{\alpha}
\def\b{\beta}
\def\c{\varepsilon}
\def\d{\delta}
\def\e{\epsilon}
\def\g{\gamma}
\def\h{\theta}
\def\k{\kappa}
\def\l{\lambda}
\def\n{\nu}
\def\p{\psi}
\def\q{\partial}
\def\r{\rho}
\def\s{\sigma}
\def\t{\tau}
\def\u{\upsilon}
\def\v{\varphi}
\def\w{\omega}
\def\D{{\mit \Delta}}
\def\G{\Gamma}
\def\H{\Theta}
\def\L{\Lambda}
\def\F{\Phi}
\def\P{\Psi}

\def\S{\Sigma}

\def\o{\over}
\def\beq{\begin{eqnarray}}
\def\eeq{\end{eqnarray}}
\newcommand{\gsim}{ \mathop{}_{\textstyle \sim}^{\textstyle >} }
\newcommand{\lsim}{ \mathop{}_{\textstyle \sim}^{\textstyle <} }
\newcommand{\vev}[1]{ \left\langle {#1} \right\rangle }
\newcommand{\bra}[1]{ \langle {#1} | }
\newcommand{\ket}[1]{ | {#1} \rangle }
\newcommand{\EV}{ {\rm eV} }
\newcommand{\KEV}{ {\rm keV} }
\newcommand{\MEV}{ {\rm MeV} }
\newcommand{\GEV}{ {\rm GeV} }
\newcommand{\TEV}{ {\rm TeV} }
\def\diag{\mathop{\rm diag}\nolimits}
\def\Spin{\mathop{\rm Spin}}
\def\SO{\mathop{\rm SO}}
\def\O{\mathop{\rm O}}
\def\SU{\mathop{\rm SU}}
\def\U{\mathop{\rm U}}
\def\Sp{\mathop{\rm Sp}}
\def\SL{\mathop{\rm SL}}

\newcommand\aap{A\&A}                
\let\astap=\aap                          
\newcommand\aapr{A\&ARv}             
\newcommand\aaps{A\&AS}              
\newcommand\actaa{Acta Astron.}      
\newcommand\afz{Afz}                 
\newcommand\aj{AJ}                   
\let\applopt=\ao                         
\newcommand\aplett{Astrophys.~Lett.} 
\newcommand\apjl{ApJ}                
\let\apjlett=\apjl                       
\newcommand\apjs{ApJS}               
\let\apjsupp=\apjs                       
\newcommand\apss{Ap\&SS}             
\newcommand\araa{ARA\&A}             
\newcommand\arep{Astron. Rep.}       
\newcommand\aspc{ASP Conf. Ser.}     
\newcommand\azh{Azh}                 
\newcommand\baas{BAAS}               
\newcommand\bac{Bull. Astron. Inst. Czechoslovakia} 
\newcommand\bain{Bull. Astron. Inst. Netherlands} 
\newcommand\caa{Chinese Astron. Astrophys.} 
\newcommand\cjaa{Chinese J.~Astron. Astrophys.} 
\newcommand\fcp{Fundamentals Cosmic Phys.}  
\newcommand\gca{Geochimica Cosmochimica Acta}   
\newcommand\grl{Geophys. Res. Lett.} 
\newcommand\iaucirc{IAU~Circ.}       
\newcommand\icarus{Icarus}           
\newcommand\japa{J.~Astrophys. Astron.} 
\newcommand\jcap{J.~Cosmology Astropart. Phys.} 
\newcommand\jgr{J.~Geophys.~Res.}    
\newcommand\jqsrt{J.~Quant. Spectrosc. Radiative Transfer} 
\newcommand\jrasc{J.~R.~Astron. Soc. Canada} 
\newcommand\memras{Mem.~RAS}         
\newcommand\memsai{Mem. Soc. Astron. Italiana} 
\newcommand\mnassa{MNASSA}           
\newcommand\mnras{MNRAS}             
\newcommand\na{New~Astron.}          
\newcommand\nar{New~Astron.~Rev.}    
\newcommand\nphysa{Nuclear Phys.~A}  
\newcommand\pasa{Publ. Astron. Soc. Australia}  
\newcommand\pasp{PASP}               
\newcommand\pasj{PASJ}               
\newcommand\physrep{Phys.~Rep.}      
\newcommand\physscr{Phys.~Scr.}      
\newcommand\planss{Planet. Space~Sci.} 
\newcommand\procspie{Proc.~SPIE}     
\newcommand\rmxaa{Rev. Mex. Astron. Astrofis.} 
\newcommand\qjras{QJRAS}             
\newcommand\sci{Science}             
\newcommand\skytel{Sky \& Telesc.}   
\newcommand\solphys{Sol.~Phys.}      
\newcommand\sovast{Soviet~Ast.}      
\newcommand\ssr{Space Sci. Rev.}     
\newcommand\zap{Z.~Astrophys.}       


\title{Constraints on mixed dark matter from anomalous \\ strong lens systems}
\author{Ayuki Kamada}
 \email{ayuki.kamada@ucr.edu}
 \affiliation{%
 Department of Physics and Astronomy, University of California, Riverside, 900 University Ave, Riverside, California 92521, USA
}%
\author{Kaiki Taro Inoue}%
 \email{kinoue@phys.kindai.ac.jp}
 \affiliation{%
 Faculty of Science and Engineering, Kindai University, Higashi-Osaka, Osaka, 577-8502, Japan 
}%
\author{Tomo Takahashi}%
 \email{tomot@cc.saga-u.ac.jp}
 \affiliation{%
 Department of Physics, Saga University, Saga 840-8502, Japan
}%

\date{\today}

\begin{abstract}
Recently it has been claimed that the warm dark matter (WDM) model cannot at the same time reproduce the observed Lyman-$\alpha$ forests in distant quasar spectra and solve the small-scale issues in the cold dark matter (CDM) model.
As an alternative candidate, it was shown that the mixed dark matter (MDM) model that consists of WDM and CDM can satisfy the constraint from Lyman-$\alpha$ forests and account for the ``missing satellite problem'' as well as the reported $3.5$\,keV anomalous X-ray line.
We investigate observational constraints on the MDM model using strong gravitational lenses.
We first develop a fitting formula for the nonlinear power spectra in the MDM model by performing $N$-body simulations and estimate the expected perturbations caused by line-of-sight structures in four quadruply lensed quasars that show anomaly in the flux ratios.
Our analysis indicates that the MDM model is compatible with the observed anomaly if the mass fraction of the warm component is smaller than 0.47 at the $95 \%$ confidence level.
The MDM explanation to the anomalous X-ray line and the small-scale issues is still viable even after this constraint is taken into account.
\end{abstract}


\maketitle

\section{Introduction}

Warm dark matter (WDM) has been investigated as a possible solution to small-scale issues for the concordant cold dark matter (CDM) model (see, e.g.,\,\citet{Bode:2000gq}).
One of such issues is known as the missing satellite problem; the predicted number of subhalos in a MW-size halo is larger than the observed one by a factor of $10$ or more \citep{Moore:1999nt, Kravtsov:2009gi}.
The thermal velocity of a DM particle suppresses the formation of such sub-galactic objects.
The suppression is usually parametrized by the thermal WDM mass.
The number count of subhalos in MW-size halos puts the lower bound on the WDM mass \citep{Polisensky:2010rw, Lovell:2013ola, Horiuchi:2013noa}.

The suppression in WDM models is not limited in subhalos in host halos, but also in field halos with a sub-galactic mass ($\sim 10^9\,\ms$).
It leads to a variety of observable implications: suppressed number of high-$z$ galaxies \citep{Pacucci:2013jfa, Schultz:2014eia, Lapi:2015zea, Menci:2016eww}, gamma-ray bursts \citep{deSouza:2013hsj, Mesinger:2005ah}, and lensed supernovae\,\citep{Pandolfi:2014rea}; delay of the reionization \citep{Barkana:2001gr, Schultz:2014eia, Lapi:2015zea},
which can be used to constrain  the WDM mass.
The WDM mass can also be constrained by observed Lyman-$\alpha$ forests in distant quasar spectra \citep{Viel:2005qj, Seljak:2006qw, Viel:2006kd, Boyarsky:2008xj, Viel:2013apy}.
The absorption lines represent the line-of-sight distribution of the neutral hydrogen and thus the underlying matter distribution.
Actually Lyman-$\alpha$ forests put the most stringent constraint on the WDM mass; $m_{\rm WDM} > 3.3$\,keV (2$\sigma$) \citep{Viel:2013apy}.
It seems difficult to evade this constraint and resolve the small-scale issues simultaneously \citep{Schneider:2013wwa}.

One minimal extension of WDM is to assume that DM consists of both cold and warm components, which is called mixed dark matter (MDM).
The Lyman-$\alpha$ forest constraint on the WDM mass in MDM models is milder than that in pure WDM models \citep{Boyarsky:2008xj}.
The structure formation in MDM models is not well established even after some previous efforts \citep{Anderhalden:2012qt, Anderhalden:2012jc}.
MDM models also attract interests in the context of the reported anomalous X-ray line in stacked X-ray spectra in
\texttt{XMM-Newton} and \texttt{Chandra} data \citep{Bulbul:2014sua, Boyarsky:2014jta}.
While the anomaly has not been confirmed in \texttt{Suzaku} data \citep{Tamura:2014mta}, the 3.5\,keV unidentified X-ray line may originate from the decay of sterile neutrinos (see, e.g., \citet{Kusenko:2009up}).
\citet{Harada:2014lma} show that decaying 7\,keV sterile neutrinos that are produced via non-resonant process called Dodelson-Widrow mechanism \citep{Dodelson:1993je} can reproduce the 3.5\,keV X-ray line if they account for $20\text{--}60\%$ of the present mass density of DM.
Interestingly this MDM model can also mitigate the missing satellite problem while evading constraints from the Lyman-$\alpha$ forests.

To constrain the clustering property of DM on (sub-)galactic scales, strong gravitational lens offers a powerful tool.
Only with a smooth gravitational potential, some quasar-galaxy lens systems with a quadruple image show a discrepancy between the observed and predicted flux ratios of multiple images.
Such a discrepancy is called the ``anomalous flux ratio'' and has been considered as an imprint of CDM subhalos with a mass of $\sim 10^{8-9} \ms$ in the lens galaxy halos \citep{mao1998, metcalf2001, chiba2002, dalal-kochanek2002, keeton2003, inoue-chiba2003, kochanek2004, metcalf2004, chiba2005, sugai2007, mckean2007, more2009, minezaki2009, xu2009, xu2010, fadely2012, macleod2013}. 

However, intergalactic halos in the line of sight can act as perturbers as well \citep{chen2003,metcalf2005a,xu2012}.
Indeed, taking into account of astrometric shift, it has been shown that the observed anomalous flux ratios can be explained solely by line-of-sight structures with a surface density $\sim 10^{7-8}\, h^{-1}\ms/\textrm{arcsec}^2$ \citep{inoue-takahashi2012, takahashi-inoue2014, inoue-etal2015, inoue-minezaki2015} without taking into account subhalos in the lens galaxies. Since the role of subhalos is relatively minor \citep{xu2015,inoue2016}, we can constrain various DM models by using the clustering property of DM in the line of sight. 

In this paper, we investigate the structure formation at $\lesssim 10$ kpc length scales in MDM models by using anomalous quadruple lenses. 
Our study is an extension of previous works for constraining pure WDM models \citep{miranda2007, inoue-etal2015}. 
To take into account non-linear clustering effects, we first calculate the non-linear power spectra of matter fluctuations down to mass scales of $\sim 10^5\, h^{-1 } \ms$ by using $N$-body simulations. 
For simplicity, we do not consider baryonic dynamics in our simulations. Then we estimate the PDF of magnification perturbation for each lens system using the semi-analytic formulae developed in \citet{takahashi-inoue2014}.

In the next section, we develop a fitting function of non-linear matter power spectra in MDM models.
This is a key input in calculation of the magnification perturbation of lensed images.
The fitting function is based on the power spectra measured from $N$-body simulations.
We provide details of our simulations setups.
In section \ref{sec:lensana}, we briefly describe our lens samples and a fiducial lens model for them.
In section \ref{sec:magperturb}, we introduce a statistic for representing the magnification perturbation of lensed images and briefly describe the semi-analytic formulation for estimating the statistic.
In section \ref{sec:result}, the magnification perturbations in the MDM models are compared to the observed values to put  constraints on the MDM models.
Section \ref{sec:concanddis} is devoted to concluding this paper and discussion on  future prospects.

Throughout this paper, we take cosmological parameters obtained from the observed cosmic microwave background (Planck + WMAP polarization, \citet{Ade:2013zuv}) to be consistent with \citet{inoue-etal2015}:
a current matter density $\Omega_{m,0}=0.3134$, a baryon density $\Omega_{b,0}=0.0487$, a cosmological constant $\Omega_{\Lambda,0}=0.6866$, a Hubble constant $H_0=67.3 (=100h)\, \textrm{km}/\textrm{s}/\textrm{Mpc}$, a spectral index $n_s=0.9603$, and the root-mean-square (rms) amplitude of matter fluctuations at $8 h^{-1}\, \textrm{Mpc}$, $\sigma_8=0.8421$.

\section{Non-linear power spectrum}
\label{sec:matterpower}

\subsection{Initial condition}
\label{subsec:inicondition}
First, we need to follow the co-evolution of the linear density fluctuations of the cold and warm components in MDM models.
To this end, we modify the public code \texttt{CAMB} suitably \citep{Lewis:1999bs}.
We assume that the warm component consists of spin-$1/2$ particles that follow the Fermi-Dirac distribution just like the conventional WDM.
In the MDM model, we have two parameters to describe its property: a mass and a temperature of the warm component $(m_{\rm WDM}$, $T_{\rm WDM})$.
Meanwhile, the resultant matter power spectra can be characterized by two parameters: the ratio of its present mass (energy) density to the whole DM density $r_{\rm warm}$ and the comoving Jeans scale at the matter-radiation equality $k_J$.
The relations among these parameters are given as follows.

The mass ratio of the warm component to the whole DM $r_{\rm warm}$ is defined as 
\BE
\label{eq:r_warm}
r_{\rm warm}  = \frac{\Omega_{{\rm warm},0} h^2}{ \Omega_{{\rm dm},0} h^2} \, ,
\EE
where $ \Omega_{{\rm dm},0}$ is the total DM mass density.
The relic mass density of the warm component is given by
\BE
\label{eq:omwarm}
\Omega_{{\rm warm},0} h^2 = r_{\rm warm} \, \Omega_{{\rm dm},0} h^2
= \left( \f{T_{\rm WDM}}{T_{\nu}} \right)^3 \left( \f{m_{\rm WDM}}{94\,{\rm eV}} \right) \,.
\EE
We assume the rest of DM consists of some stable and cold particles such that $\Omega_{{\rm cold},0} + \Omega_{{\rm warm},0} = \Omega_{{\rm dm},0}$.

The comoving Jeans scale at the matter-radiation equality $t = t_{\rm eq}$ is given by \citep{Kamada:2013sh},
\BEA
\label{eq:kJ}
k_{\rm J} &=& a \sqrt{\f{4 \pi G \rho_{\rm M}}{\sigma^2}} \bigg|_{t=t_{\rm eq}} \N \\
&=& 14/{\rm Mpc} \, \left( \f{m_{\rm WDM}}{0.5\,{\rm keV}} \right)^{4/3} \left( \f{0.5}{r_{\rm warm}} \right)^{5/6} \,,
\EEA
where $a$ is the scale factor of the Universe, $G$ is the gravitational constant, $\rho_{\rm M}$ is the matter mass density, and $\sigma^2$ is the mass-weighted mean squared velocity of the whole DM.
In the MDM model, $\sigma^2$ is the sum of two contributions $\sigma_{\rm cold}^2$ + $\sigma_{\rm warm}^2$.
The former is $\sigma_{\rm cold}^2 = 0$ by definition and the latter is $\sigma_{\rm warm}^2 = r_{\rm warm} \, \sigma_{\rm WDM}^2$, where $\sigma_{\rm WDM}^2$ is the mean squared velocity of the warm component.
In the second equality of eq. (\ref{eq:kJ}), we use eq. (\ref{eq:omwarm}) to eliminate $T_{\rm WDM}$.

After checking that our modified version of \texttt{CAMB} reproduces the results in \citet{inoue-etal2015} in the pure CDM and WDM limits, we calculate the resultant linear matter power spectra in 17 models listed in table \ref{tab:modelparam}.
Some of the models are the same as in \citet{Anderhalden:2012jc}.
We show some of the linear matter power spectra that are extrapolated to the present $z=0$ in figure \ref{fig:linpower}.
The suppression at $k \gg k_{\rm J}$ is milder for smaller $r_{\rm warm}$.
We use the linear matter power spectra to generate the initial condition of $N$-body simulation, which we discuss in the following.

\begin{table}
\begin{tabular}{lcc}
\hline \hline
Model & $m_{\rm WDM}$ [keV] & $r_{\rm warm}$ \\
\hline
CDM & - & 0 \\
\hline
MDM(0.02, 0.05) & 0.02 & 0.05 \\
MDM(0.05, 0.05)$^{\rm F}$  & 0.05 & 0.05 \\
MDM(0.1, 0.05)$^{\rm F}$& 0.1 & 0.05 \\
MDM(0.1, 0.1) & 0.1 & 0.1 \\
MDM(0.1775, 0.1) & 0.1775 & 0.1 \\
MDM(0.1, 0.2)$^{\rm F}$ & 0.1 & 0.2 \\
MDM(0.3, 0.2)$^{\rm F}$ & 0.3 & 0.2 \\
MDM(0.25, 0.4) & 0.25 & 0.4 \\
MDM(0.4, 0.4) & 0.4 & 0.4 \\
MDM(0.525, 0.4) & 0.525 & 0.4 \\    
MDM(0.3, 0.5)$^{\rm F}$ & 0.3 & 0.5 \\  
MDM(0.5, 0.5) & 0.5 & 0.5 \\  
MDM(0.757, 0.6) & 0.757 & 0.6 \\   
MDM(1, 0.8)$^{\rm F}$ & 1 & 0.8 \\  
\hline 
WDM-1.3 & 1.3 & 1 \\   
WDM-2 & 2 & 1 \\ 
\hline
\end{tabular}
\caption{\label{tab:modelparam} Simulated models. The models denoted by $^{\rm F}$ are used for obtaining the fitting formula presented in \ref{subsec:sim}.}
\end{table}

\BF
\IG[width=85mm]{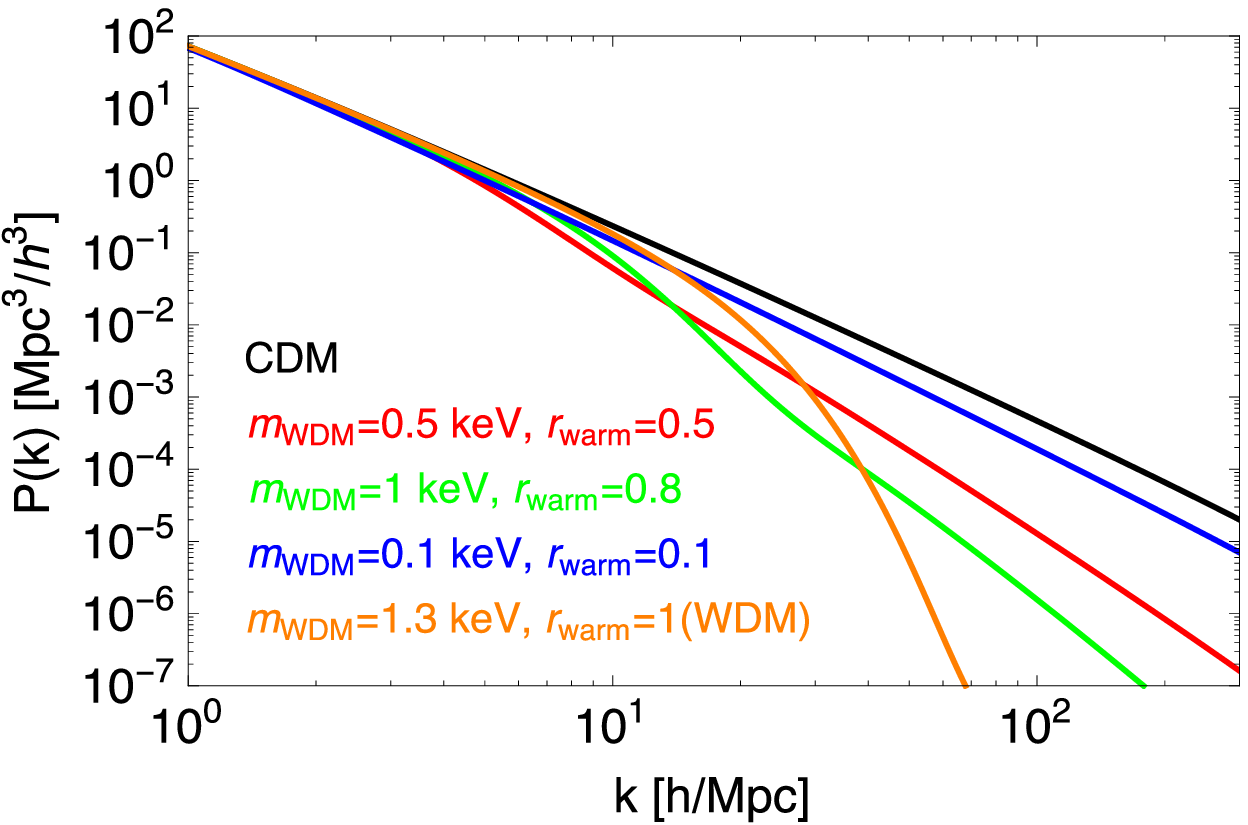}
\caption{Linear matter power spectra at present. Among the models in table \ref{tab:modelparam},
we compare models of CDM (black), MDM(0.5, 0.5) (red), MDM(1, 0.8) (green), MDM(0.1, 0.1) (blue) and WDM-1.3 (orange).
We note that WDM-1.3 has been excluded with $95$\% confidence level by using the same four anomalous samples of quadruple lenses as in this paper \citet{inoue-etal2015}.
}
\label{fig:linpower}
\EF

\subsection{$N$-body simulation}
\label{subsec:sim}
We perform $N$-body simulations by using the public code \texttt{Gadget-2} \citep{Springel:2005mi}.
Our simulation setups (L5, L10, HL10) are summarized in table \ref{tab:simset}.
We initiate all the simulations from $z=49$.
We measure the matter power spectra from the simulated matter distributions at $z=0,0.3,0.6,1,2$, and $3$.

\begin{table}
\begin{tabular}{lccc}
\hline \hline
Setup & $L\,[{\rm Mpc}/h]$ & $N$ & $\epsilon\,[{\rm kpc}/h]$ \\
\hline
L5 & 5 & $512^{3}$ & 0.5 \\
\hline
L10 & 5 & $512^3$ & 1.0 \\
\hline 
HL10 & 10 & $1024^3$ & 0.5 \\
\hline
\end{tabular}
\caption{\label{tab:simset} Simulation setups. 
$L$ is the length on a side of the simulation box, $N$ is the number of the simulation particles, and $\epsilon$ is the gravitational softening length. 
HL10 is available only in CDM and MDM(0.5, 0.5), which are discussed in appendix \ref{sec:L10HL10}.}
\end{table}

We obtain the fitting function of the measured non-linear matter power spectra through the following steps.
First we run L5 simulations in six models denoted by $^F$ in table \ref{tab:modelparam}.
We assume the fitting function takes a form of
\BE
\label{eq:transfer}
P_{\rm MDM}/P_{\rm CDM} =  T^{2}(f, k'_{\rm d}) = (1-f_{\rm warm}) + \f{f_{\rm warm}}{(1+k/k'_{\rm d})^{0.7441}} \,,
\EE
with functions of
\BEA
f_{\rm warm} (r_{\rm warm}) &=& 1-\exp \left( - a \f{r_{\rm warm}^{b}}{1-r_{\rm warm}^{c}} \right) \,, \\
k'_{\rm d} (k_{\rm d}, r_{\rm warm}) &=& k_{\rm d} / r_{\rm warm}^{5/6} \,,
\EEA
where $a$, $b$, and $c$ are positive parameters and
\BE
k_{\rm d} (m_{\rm WDM}, z) = 388.8 \, h/{\rm Mpc} \left( \f{m_{\rm WDM}}{\rm keV} \right)^{2.207} D(z)^{1.583}\,,
\EE
is the damping scale given in \citet{inoue-etal2015} with the linear growth rate $D(z)~(D(0)=1)$.
The dependence of $k'_{\rm d}$ on $r_{\rm warm}$ ($k'_{\rm d} \propto r_{\rm warm}^{-5/6}$) is inferred by that of $k_{\rm J}$ in eq. (\ref{eq:kJ}).
Here we remark that $f_{\rm warm}(r_{\rm warm}=0)=0$ and $f_{\rm warm}(r_{\rm warm}=1)=1$ and thus the above fitting function reproduces the results in \citet{inoue-etal2015} in the pure CDM and WDM limits.

Next we determine the three parameters $a$, $b$, and $c$ by minimizing the residual of 
\BE
\sum_{\rm MDM \,  models} \sum_{z \, {\rm bins}} \sum_{k \, {\rm bins}} \left( T^{2} - P_{\rm MDM}/P_{\rm CDM} |_{\rm L5} \right)^2 \,,
\EE
where MDM models are those denoted by $^{\rm F}$ in table \ref{tab:modelparam},
$z \, {\rm bins}$ are $z \in \{0,0.3,0.6,1,2,3\}$, and $k \, {\rm bins}$ are $\log(k \, [h/\,{\rm Mpc}]) \in \{2.117,2.137,2.157,\cdots,2.477\}$ (total 20 bins).
Finally we find $a=1.551$, $b=0.5761$, and $c=1.263$.

We compare the fitting function of $T^2$ with simulated $P_{\rm MDM}/P_{\rm CDM} |_{\rm L5}$ in figure \ref{fig:transfer}.
As a check, we also perform L10 simulations to measure the matter power spectra.
The L10 simulations have a larger box and a lower mass resolution in comparison with the L5 simulation (see table \ref{tab:simset}).
As discussed in appendix \ref{sec:L10HL10}, the L10 simulation is concordant with the HL10 simulation that has a larger box and equivalent mass resolution in comparison with the L5 simulation.
The L10 simulation generically shows at most $10$\% larger power spectra than those from the L5 simulation (see figure \ref{fig:transfer}).
This is possibly due to the relatively small box size of our simulations.
It allows us to simulate the matter distribution at the relevant scale ($k_{\rm lens} \sim 300\,h/$Mpc, see section \ref{sec:magperturb}), while it may miss ($\lesssim 10$\%) contributions to power spectra from larger-scale structure formation.

\BF
\IG[width=85mm]{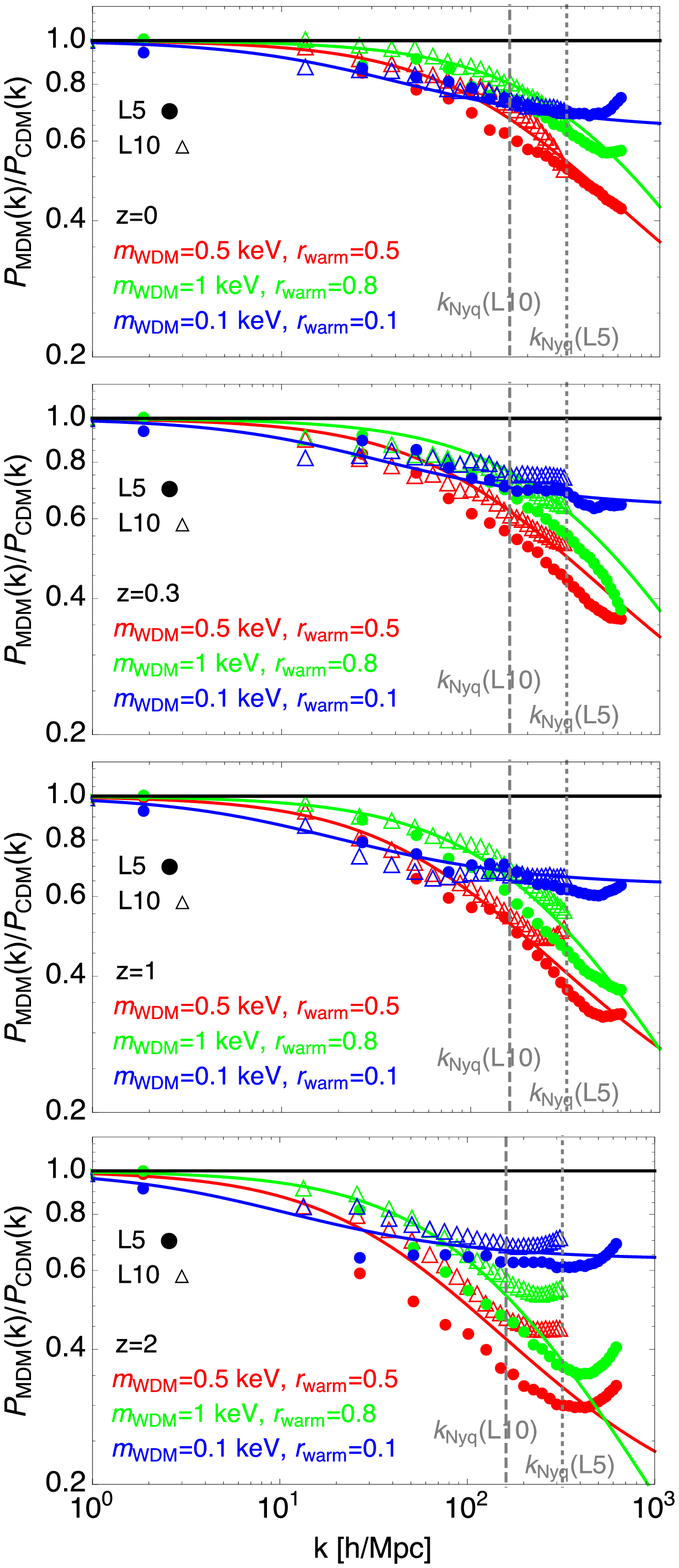}
\caption{Comparison of $T^{2}$ with $a=1.551$, $b=0.5761$, and $c=1.263$ and $P_{\rm MDM}/P_{\rm CDM} |_{\rm L5, L10}$ at $z=0,0.3,1,$ and $2$. 
We take the same MDM models as in figure \ref{fig:linpower}. 
MDM(1, 0.8) (green) with the other five MDM models are used to determine the parameters of the fitting function $T^{2}$. 
This also gives a reasonable fit up to $20$\% to others including a model with MDM(0.5, 0.5) (red) and MDM(0.1, 0.1) (blue) that are not used in the calibration of the fitting function.
We present the Nyquist wavenumbers of the L5, L10, and HL10 simulations (dashed lines), above which the measured matter power spectra are reliable.}
\label{fig:transfer}
\EF

As a further check, we perform the L5 and L10 simulations in additional eight MDM models that are not used in the calibration of the fitting function.
Importantly, we find that our fitting function reproduces $P_{\rm MDM}/P_{\rm CDM} |_{\rm L5}$ within $20$\% even in theses additional MDM models.
We use the \texttt{halofit} for $P_{\rm CDM}$ that is calibrated with the same cosmological parameters as employed in this paper.
The explicit expression can be found in \citet{takahashi-inoue2014, inoue-etal2015}, and thus not repeated here.
Now we can calculate $P_{\rm MDM}$ by combining eq. (\ref{eq:transfer}) and the \texttt{halofit}.
We find that the calculated $P_{\rm MDM}$ overpredicts $P_{\rm MDM} |_{\rm L5(L10)}$ by at most $\sim 40\% (20\%)$ as shown in figure \ref{fig:nonlinear}.
We remark that this is conservative when putting constraints on the $(m_{\rm WDM}, r_{\rm warm})$-plane (see section\,\ref{sec:result}), while the resultant $\langle \eta^2 \rangle^{1/2}$ is enlarged by $\sim 10\%$ or less where $\langle \eta^2 \rangle^{1/2}$ represents a magnification perturbation and is explained in section \ref{sec:magperturb}.

\BF
\IG[width=85mm]{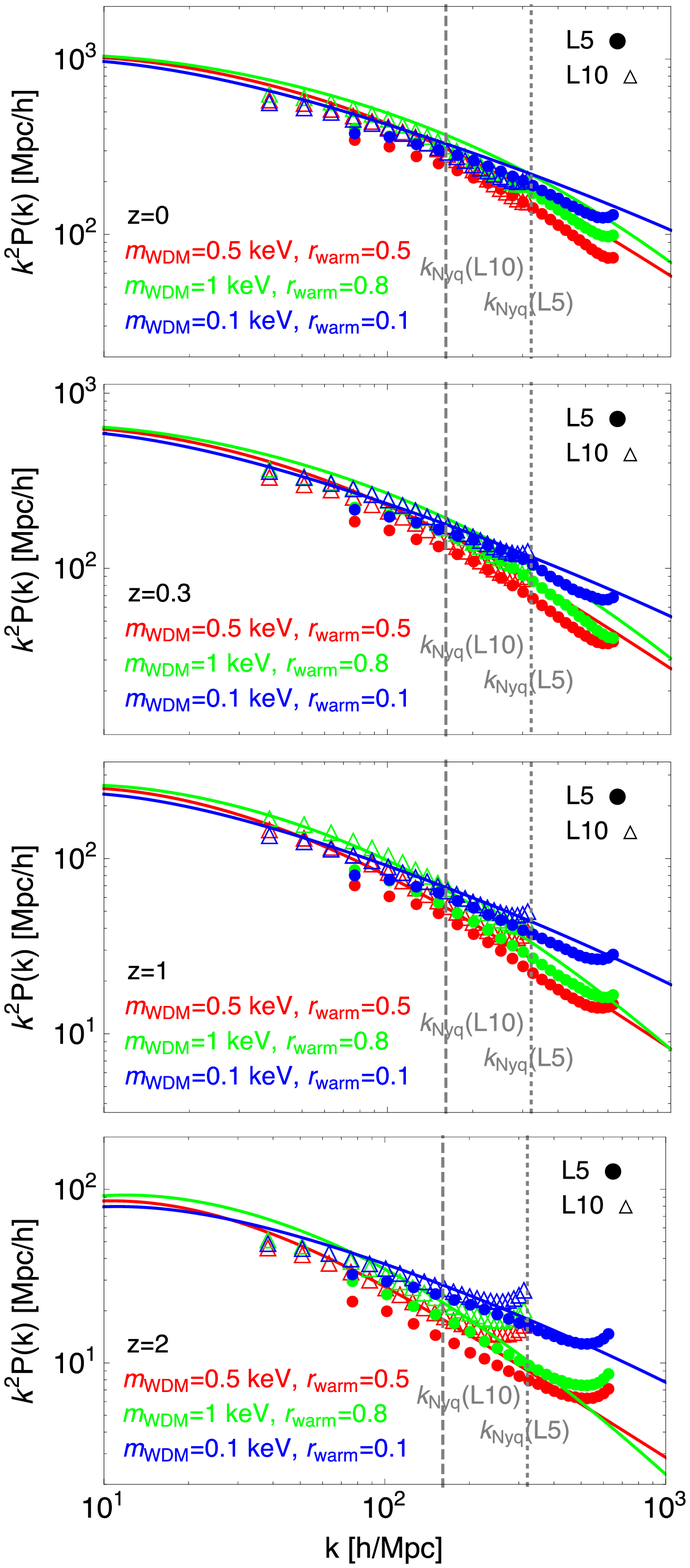}
\caption{Comparison of $P_{\rm MDM}$ (\texttt{halofit} multiplied by $T^{2}$) and $P_{\rm MDM}|_{\rm L5, L10}$ at $z=0,0.3,1,$ and $2$. We take the same MDM models as in figure \ref{fig:linpower}.}
\label{fig:nonlinear}
\EF

\section{Lens analysis}
\label{sec:lensana}

\subsection{Systems}
\label{subsec:sys}

In what follows, we use four anomalous quadruple lenses B1422+231, B0128+437, MG0414+0534, and B0712+472 with source redshifts $1<z_{\tr{S}}<4$.
The flux ratios in these systems show more than $2\sigma$ deviation in comparison with the prediction for a best-fitted smooth lens model described in the next section. 
The data used in our analysis are listed in table \ref{tab:data}.
For details, we refer the readers to \citet{inoue-etal2015}.

\begin{table*}
\hspace{-1.5cm}
\begin{tabular}{lccccccccccr}
\hline
\hline
lens system  & $z_{\tr{S}}$ & $z_{\tr{L}}$ &position& flux  & $N_{\tr{image}}$ &  $b$($''$) & $\langle \kappa \rangle $ & $k_{\tr{lens}}(h/\tr{Mpc})$ & $R_E$(kpc) 
& $\hat{\eta}$ & references   \\ 
\hline
B1422+231 & 3.62 & 0.34  & opt/NIR & radio & $3$ & $0.78$ & $0.40$ & $412$ & $3.9$ 
& $0.098\pm0.005$ & (1) (2)
\\
\hline
B0128+437 & 3.124 & 1.145  & opt/NIR & radio & $4$ & $0.24$  & $0.52$ & $527$ & $2.1$ 
& $0.0632\pm 0.025$ & (1) (3) (4)
\\
\hline
MG0414+0534 & 2.639 & 0.96  & opt/NIR & MIR & $4$ & $1.1$ & $0.55$ &  $118$ & $9.0$ 
& $0.131\pm 0.042$ & (5) (6) (7)
\\
\hline
B0712+472 & 1.339 & 0.406  & opt/NIR & radio & $3$ & $0.77$ & $0.50$ & $401$ & $4.3$ 
& $0.131\pm 0.071$ & (1) (5)
\\
\hline
\end{tabular}
\caption{\label{tab:data} Anomalous quadruple lens systems used in our analysis. $z_{\tr{S}}$ and $z_{\tr{L}}$ are the source and lens redshifts. 
$\langle \kappa \rangle $ is the convergence of the best-fitted model averaged over those at the positions of $N_{\tr{image}}$ lensed images. 
$b$ is the mean angular separation between a lensed image and a lens centre. 
$R_{E}$ is the effective Einstein radius in the proper coordinates. 
$\hat{\eta}$ is the observed $\eta$ for the best-fitted model. 
References: (1) \citet{koopmans2003} (2) \citet{sluse2012}  (3) \citet{biggs2004} (4) \citet{lagattuta2010} (5) CASTLES data base:http://www.cfa.harvard.edu/castles (6) \citet{minezaki2009}    (7) \citet{macleod2013}}
\end{table*}

\subsection{Model}
\label{subsec:model}

For modeling a primary lens galaxy halo, we use a singular isothermal ellipsoid (SIE) \citep{kormann1994}. 
The parameters of SIE are the effective Einstein radius, which corresponds to the mass scale inside a critical curve, the ellipticity and the position angle of the lens, and the positions of the lens centre and the source. 
To take into account the effects of groups, clusters, and large-scale structures to the primary lens at angular scales larger than the effective Einstein radius, an external shear is also included in our analysis. 
The strength and direction of shear are also used as fitting parameters.
For our analysis, we use a public code  \texttt{GRAVLENS}\footnote{http://redfive.rutgers.edu/$\sim$keeton/gravlens/}.
For other details, the readers are referred to  \citet{inoue-etal2015}.


\section{Magnification perturbation}
\label{sec:magperturb}

In this section, we briefly describe our method for evaluating the magnification perturbation induced by the line-of-sight structure. 
For details, we refer the readers to \citet{inoue-etal2015}, in which the same method is used as that given in \citet{inoue-takahashi2012,takahashi-inoue2014}.

To characterize the strength of perturbation in the magnification of lensed images in strong lens systems, we adopt a statistic $\eta$, which represents the expected magnification perturbation per lensed image.  
Suppose that multiple images of a point source consist of $N_{\rm pair}$ pairs of images with different parities.  
Then $\eta$ can be written in terms of magnifications $\mu_i$ of the unperturbed lens and their perturbations $\delta \mu_i$ at the positions of multiple ``$i$'' images,
\BE
\eta \equiv \biggl[\frac{1}{2 N_{\rm pair}} \sum_{ i \neq j} \left[ \delta^\mu_{i} ({\rm minimum}) - \delta^\mu_{ j} ({\rm saddle}) \right]^2 \biggr]^{1/2},
\label{eta_def}
\EE
where $\delta^\mu_i \equiv \delta \mu_i/\mu_i$ represents a magnification contrast of ``$i$'' image and ``minimum'' and ``saddle'' correspond to a minimum and a saddle points in the arrival time surface. 

Assuming that the strong lens effect from perturbers is negligible, the perturbed lens equation at a certain lens plane is given by
\BE
\tilde{{\bm \theta}}_y = \left( {\bm 1} - {\bm \Gamma}_i - \delta {\bm \Gamma}_i \right) {\bm \theta}_x,
\EE
where ${\bm \theta}_x$ and $\tilde{{\bm \theta}}_y$ are the angular position of the lensed image and that of the source image, respectively. 
${\bm \Gamma}_i$ and $\delta {\bm \Gamma}_i$ can be written with the convergence $\kappa_i$ and the shear components $\gamma_{i1}, \gamma_{i2}$ and their perturbations $\delta \kappa_i, \delta \gamma_{i1},$ and $\delta \gamma_{i2}$ at a lens plane as 
\BE 
{\bm \Gamma}_i + \delta {\bm \Gamma}_i =
\begin{pmatrix}
\kappa_i + \gamma_{i1} & \gamma_{i2}   \\
\gamma_{i2}  &   \kappa_i - \gamma_{i1}
\end{pmatrix}
+
\begin{pmatrix}
\delta  \kappa_i + \delta\gamma_{i1} & \delta\gamma_{i2}   \\
\delta\gamma_{i2}      &  \delta \kappa_i - \delta\gamma_{i1}
\end{pmatrix},
\EE
from which the perturbed magnification is given by
\BEA
(\mu_i + \delta \mu_i)^{-1} &=& ( 1 - \kappa_i - \gamma_{i1} -\delta \kappa_{i} - \delta \gamma_1) ( 1 - \kappa_i + \gamma_{i1} -\delta \kappa_{i} + \delta \gamma_1) - (\gamma_{i2} + \delta \gamma_{i2} )^2.
\EEA
Up to the linear order, the magnification contrast can be obtained by
integrating the convergence and shear perturbations along the photon path, 
\BE
\delta_i^\mu = \int_{\tr{photon path}} \frac{2(1-\kappa_i) \delta \kappa_{i} + 2 \gamma_{i1} \delta \gamma_{i1} + 2 \gamma_{i2} \delta \gamma_{i2}}{ (1-\kappa_i)^2 - \gamma_{i1}^2 - \gamma_{i2}^2}.
\label{eq:delta_mu}
\EE

To show how we calculate the expected $\eta$ in a more explicit manner, we discuss the case with three images with two minima A, C and one
saddle B in which the separation angles between the images of A and B and those of B and C are 
$\theta_{\tr{A}\tr{B}}$ and $\theta_{\tr{B}\tr{C}}$, respectively.
In this case, the second moment of the statistic $\eta$ is given by
\BE
\eta^2 = \frac14 \left[ \left( \delta_A^\mu - \delta_B^\mu \right)^2 + \left( \delta_C^\mu - \delta_B^\mu \right)^2 \right].
\EE
In terms of observed fluxes ${\tilde \mu}_{\tr{A}}, {\tilde \mu}_{\tr{B}}, {\tilde \mu}_{\tr{C}}$ and unperturbed fluxes ${\hat \mu}_{\tr{A}}, {\hat \mu}_{\tr{B}},{\hat \mu}_{\tr{C}}$ from a best-fitted model, the estimator is given by
\BE
\hat{\eta}^2 = \f{1}{4}\biggl[\biggl( \f{{\tilde \mu}_{\tr{A}} {\hat \mu}_{\tr{B}}}{{\hat \mu}_{\tr{A}}  {\tilde \mu}_{\tr{B}}} - 1 \biggr)^2
+ \biggl( \f{{\tilde \mu}_{\tr{C}} {\hat \mu}_{\tr{B}}}{{\hat \mu}_{\tr{C}} {\tilde \mu}_{\tr{B}}} - 1 \biggr)^2 \biggr].
\EE
The ensemble average of $\eta^2$ can be estimated as follows.
Since the magnification contrast $\delta_i^{\mu}$ is given by eq. \eqref{eq:delta_mu}, in coordinates where $\langle \delta \kappa \delta \gamma_2 \rangle$ and $\langle \delta \gamma_1 \delta \gamma_2 \rangle$ vanish, the ensemble average of $\eta^2$ is 
\BEA
\langle \eta^2 \rangle &=& \frac{1}{4}\biggl[(I_A+I_B)-2I_{AB}(\theta_{AB})+(I_B+I_C) - 2I_{BC}(\theta_{BC}) \biggr],
\label{eq:estimator-anal}
\EEA
where 
\BE
I_i\equiv \mu_i^2(4(1-\kappa_i)^2+2 \gamma_{1i}^2 + 2\gamma_{2i}^2) \langle \delta \kappa (0) \delta \kappa(0) \rangle,
\label{eq:Ii}
\EE
and
\BEA
I_{ij}(\theta_{ij}) &\equiv& 4 \mu_i \mu_j \biggl[(1-\kappa_i)(1-\kappa_j) \langle \delta \kappa (0) \delta \kappa(\theta_{ij}) \rangle 
+ \gamma_{1i}\gamma_{1j} \langle \delta \gamma_1 (0) \delta \gamma_1 (\theta_{ij}) \rangle +\gamma_{2i}\gamma_{2j} \langle \delta \gamma_2 (0) \delta \gamma_2(\theta_{ij}) \rangle \nonumber \\
&& + (1-\kappa_i)\gamma_{1j} \langle \delta \kappa_i(0) \delta \gamma_{1j} (\theta_{ij}) \rangle
+ (1-\kappa_j)\gamma_{1i} \langle \delta \kappa_j(0) \delta \gamma_{1i}(\theta_{ij}) \rangle
\biggr].
\label{eq:Iij}
\EEA
We assume that fluctuations with comoving length scales larger than that of the mean separation between a lensed image and a lens centre are taken into account as components in the unperturbed lens. 
Then the correlation of convergence between a pair of points separated by $\theta$ is approximately given by
\BEA
\langle \delta \kappa (0) \delta \kappa (\theta) \rangle &=& \frac{9 H_0^4 \Omega_{m,0}^2}{4 c^4} \int_0^{r_S} dr  r^2 \biggl(\frac{r-r_S}{r_S} \biggr)^2 [1+z(r)]^2 \nonumber \\
&& \times \int_{k_{\rm {lens}}}^{k_{\tr{max}}}\frac{dk}{2 \pi} k W_{\textrm{CS}}^2(k;k_{\textrm{cut}})  P_{\delta}(k,r) J_0(g(r) k\theta),  \nonumber \\
\label{eq:ck}
\EEA
where $P_\delta (k,r)$ is the non-linear power spectrum at the comoving distance $r$ from the observer along the photon path, $r_s$ is the comoving distance from the observer to the source, $z(r)$ is the redshift to the comoving distance $r$ along the photon path, and
\BE
g(r)= \left\{ 
\begin{array}{ll}
r, & \mbox{$r<r_L$} \\
{r_L(r_S-r)}/{(r_S-r_L)}, & \mbox{$r\ge r_L$},
\end{array}
\right.
\label{eq:g}
\EE
with $r_L$ being the comoving distance to the lens galaxy and $k_{\tr{lens}}\equiv \pi/ (2 r_{L} b)$. 
Here $b$ is the mean angular separation between a lensed image and a lens centre.
$J_0$ is the 0th order Bessel function and should be replaced by $(J_0 + J_4)/2$, $(J_0 - J_4)/2$, and $- J_2$ for $\langle \delta \gamma_1(0) \delta \gamma_1(\theta) \rangle$, $\langle \delta \gamma_2(0) \delta \gamma_2(\theta) \rangle$, and $\langle \delta \kappa(0) \delta \gamma_1(\theta) \rangle$, respectively, where $J_2$ and $J_4$ are the 2nd and 4th order Bessel functions, respectively.
$W_{\rm CS}$ is the constant shift (CS) filter whose explicit expression can be found in \citet{takahashi-inoue2014}.
Through $W_{\rm CS}$, fluctuations below $k_{\tr{cut}}$ are only partially taken into account.
$k_{\tr{cut}}$ is determined such that the perturbation in relative angular positions of lenses does not exceed the maximum error in those of fitted lensed images.
$k_{\tr{max}}$ corresponds to the scale above which perturbations become negligible due to the finite source size.

From dust reverberation, the radius of the mid-infrared emitting region of MG0414+0534 is estimated as $r_\tr{s}\sim 2\,$pc \citep{minezaki2009}, which gives $k_{\tr{max}}= \pi/(2 r_s)\sim 8\times 10^4\, h/\tr{Mpc}$.   
For radio sources, we can estimate $k_{\tr{max}}$ from the apparent angular sizes of lensed VLBI images. 
Then we find that $3\times 10^3 \,h/\tr{Mpc} \lesssim k_{\tr{max}} \lesssim 1\times 10^5\,h/\tr{Mpc}$.  
In this analysis, taking into account ambiguity in the source size, we adopt a constant cut-off $k_{\tr{max}}=10^4\,h/\tr{Mpc}$.

For a given $\langle \eta^2 \rangle$, we assume the following probability density function (PDF) for $\eta$,
\BE
P(\eta) \propto \exp\left[- \left\{ \ln(1+\eta/\eta_{0}) -\ln(\mu) \right\} / (2\sigma^2) \right] / (\eta + \eta_{0}) \,,
\label{eq:PDF}
\EE
where three parameters $\eta_{0} (\langle \eta^{2} \rangle^{1/2})$,
$\mu$, and $\sigma^2$ are calibrated by ray-tracing simulations in the CDM model \citep{takahashi-inoue2014} such that
\BE
\eta_{0} (\langle \eta^{2} \rangle^{1/2}) = 0.228 \langle \eta^{2} \rangle^{1/2},~ \mu=4.10,~ \sigma^2 = 0.279 \,.
\label{eq:PDFpara}
\EE
In what follows, we use the same PDF in the MDM and WDM models as in the CDM model except for $\langle \eta^{2} \rangle^{1/2}$ replaced in each model.
  
By using the fitting formula for the non-linear power spectrum in the MDM model provided in the previous section, one can compare the model with observations of anomalous quadruple lenses listed in table \ref{tab:data}, from which we can obtain constraints on the fraction of the warm component in the total DM and the WDM mass.

\section{Results}
\label{sec:result}

We generate matter power spectra (\texttt{halofit} multiplied by the fitting function) in $48$ models including the MDM and WDM (total $16$) models listed in table \ref{tab:modelparam}.
By using them, we calculate the square root of the second moment $\langle \eta_{i}^{2} \rangle^{1/2} (m_{\rm WDM}, r_{\rm warm})$ for each lens system $i$ as explained in the previous section.
We evaluate $p$-value, which is given by
\BEA
p (m_{\rm WDM}, r_{\rm warm}) = \left( \int_{{\hat V}} \prod_{i} d \eta_{i} P (\eta_{i}; \langle \eta_{i}^{2} \rangle^{1/2}, \delta {\hat \eta}_{i})\right) {\bigg /} \left( \int \prod_{i} d \eta_{i} P (\eta_{i}; \langle \eta_{i}^{2} \rangle^{1/2}, \delta {\hat \eta}_{i}) \right) \,,
\EEA
where the PDF is integrated over all the values of $\eta_{i} \in (0, \infty) $ in the denominator, while over the following limited domain in the numerator,
\BE
\prod_{i} P (\eta_{i} \in {\hat V}; \langle \eta_{i}^{2} \rangle^{1/2}, \delta {\hat \eta}_{i}) < \prod_{i} P ({\hat \eta}_{i}; \langle \eta_{i}^{2} \rangle^{1/2}, \delta {\hat \eta}_{i}) \,.
\EE
The observational error $\delta {\hat \eta}_{i}$ is incorporated by the replacement of $\langle \eta_{i}^{2} \rangle^{1/2} \to (\langle \eta_{i}^{2} \rangle + \delta {\hat \eta}_{i}^{2})^{1/2}$ in the PDF (eqs. \eqref{eq:PDF} and \eqref{eq:PDFpara}).
The $p$-value represents the probability of finding a sample of $\eta_{i}$ that is more unlikely than the observed value of ${\hat \eta}_{i}$.

We interpolate the resultant $p$-values linearly and show the result in figure \ref{fig:pmdm}.
The constraint on $m_{\rm WDM}$ becomes weaker for smaller $r_{\rm warm}$.
This is because for a given $m_{\rm WDM}$, the suppression of linear matter power spectra are milder for smaller $r_{\rm warm}$.
MDM models with $r_{\rm warm}<0.47$ are compatible since $p>0.05$.

\BF
\IG[width=85mm]{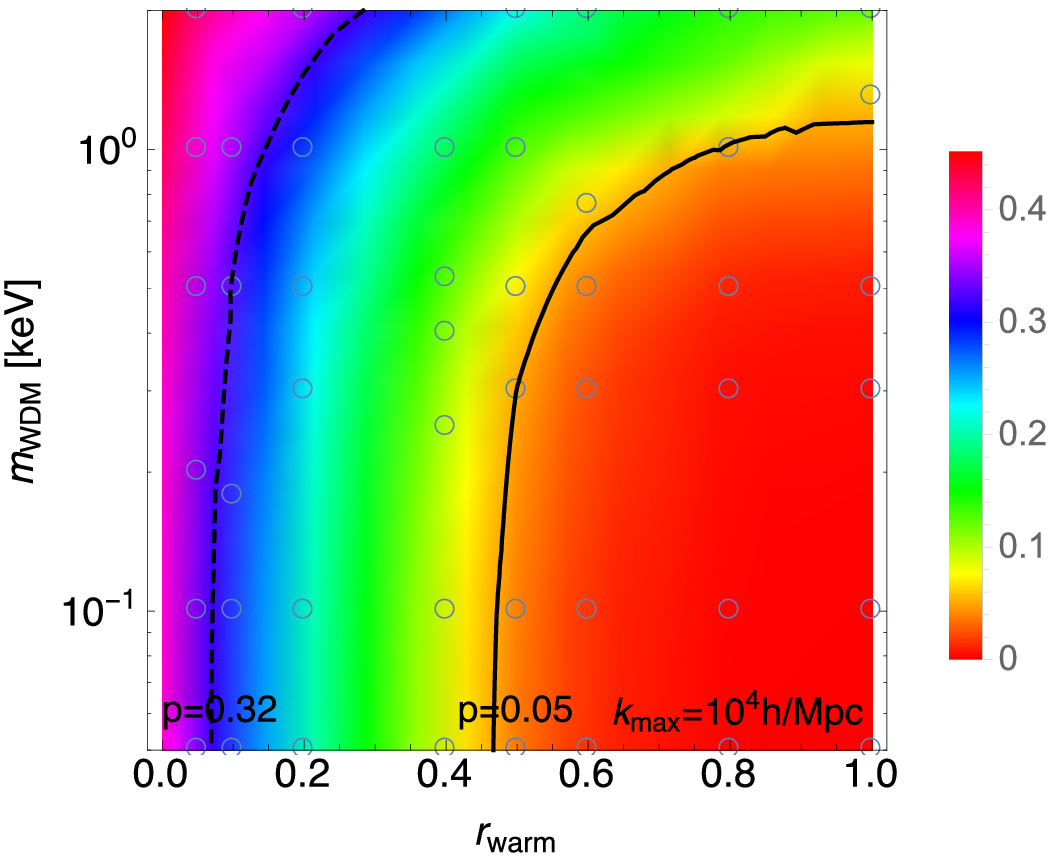}
\caption{$p$-value as a function of $m_{\rm WDM}$ and $r_{\rm warm}$. We set a cut-off scale $k_{\rm max}=10^4\,h/$Mpc. 
Circles denote the parameter sets where we evaluate $p$-values directly through the \texttt{halofit} multiplied by the fitting formula. 
These $p$-values are interpolated linearly to those at other parameter sets.
}
\label{fig:pmdm}
\EF

\section{Conclusion and Discussion}
\label{sec:concanddis}

We investigated the lensing effects of line-of-sight structures in the MDM model where DM consists of both cold and warm components.  
We have used quadruply lensed systems that show anomaly in the flux ratios in lensed images.
We have extended the previous works in the CDM and WDM models \citep{inoue-takahashi2012, inoue-etal2015} to 
include the MDM model by using the same semi-analytic formulation.
A key input in the formulation is a non-linear matter power spectrum.
We have developed a fitting function of the non-linear matter power spectra measured from the $N$-body simulations.
The fitting function reproduces the \texttt{halofit} in the pure CDM limit and that calibrated in \citet{inoue-etal2015}
in the pure WDM limit.

We examined if MDM models can account for the anomalies in flux ratios.
We confirmed that CDM and WDM models with $m_{\rm WDM} > 1.3$\, keV can be concordant with the observations.
If the mass fraction of the warm component is smaller than 0.47, then all the MDM models are compatible with $95\%$ confidence level or more.
In some models that we examined, the fitting function appears to overpredict the simulated matter power spectra up to $\sim 40 \%$.
It means that the obtained constraint is conservative, while simulations with a larger boxsize and a finer resolution are needed to reach a definite conclusion. 
Our result is compatible with the previous analysis of Lyman-$\alpha$ forest data \citep{Boyarsky:2008xj}.
They report that the data allow any value of the WDM mass if $r_{\rm warm} < 0.35$.
The MDM model with 7\,keV sterile neutrinos being warm components and $r_{\rm warm} = 0.2\text{--} 0.6$ not only explain the 3.5\,keV X-ray line \citep{Harada:2014lma} but also satisfy constrains obtained in this paper.
Interestingly, the MDM explanation to the anomalous X-ray line and the small-scale issues is still viable.
Further testing of the MDM explanation can be done with an increased number of lens samples in the near future.

In our calculations, we assume that PDF in MDM and WDM models are the same as in the CDM model except for $\langle \eta_{i}^{2} \rangle^{1/2}$. 
In order to verify this assumption, we need to perform ray-tracing Monte Carlo simulations, which will be carried out in our future work.
In our simulations, we did not take into account non-luminous subhalos hosted by the lensing galaxy halos. In the CDM model, it has been shown that the contribution of magnification perturbation caused by subhalos is less than $\sim 30\%$  for lens systems with a source redshift $z_{\tr{S}}>2.0$ \citep{inoue2016}.
As the number density of subhalos with sizes that are comparable to or less than the free-streaming length $\sim 1/k_{\tr{J}}$ is significantly reduced, the role of dark subhalos in MDM models would be subdominant. 
However, we may need to check the lensing effects caused by subhalos in MDM models as well.

\section*{Acknowledgments}
\label{sec:acknow}
This work is supported in part by JSPS Grant-in-Aid for
Scientific Research (B) (No. 25287062) ``Probing the origin
of primordial minihalos via gravitational lensing phenomena''.
The work of TT  is partially supported by JSPS KAKENHI Grant Number 15K05084  and 
MEXT KAKENHI Grant Number 15H05888.
Numerical computations were carried out on Cray XC30 at Center for Computational Astrophysics, 
National Astronomical Observatory of Japan and on SR16000 at YITP in Kyoto University.

\appendix

\section{Large box simulations}
\label{sec:L10HL10}

In this appendix, we compare the L10 simulation with the higher resolution simulation (HL10).
As shown in figure \ref{fig:nonlinearH}, two simulations are concordant at the level of $10$\% or less.
They are also close to the L5 simulation on most scales.
These results allow us to use L5 simulations to determine the parameters of the fitting function as long as we are tolerant of the $40$\%
difference between the $P_{\rm MDM}$ (\texttt{halofit} multiplied by $T^{2}$) and $P_{\rm MDM}|_{\rm L5}$.

\BF
\IG[width=85mm]{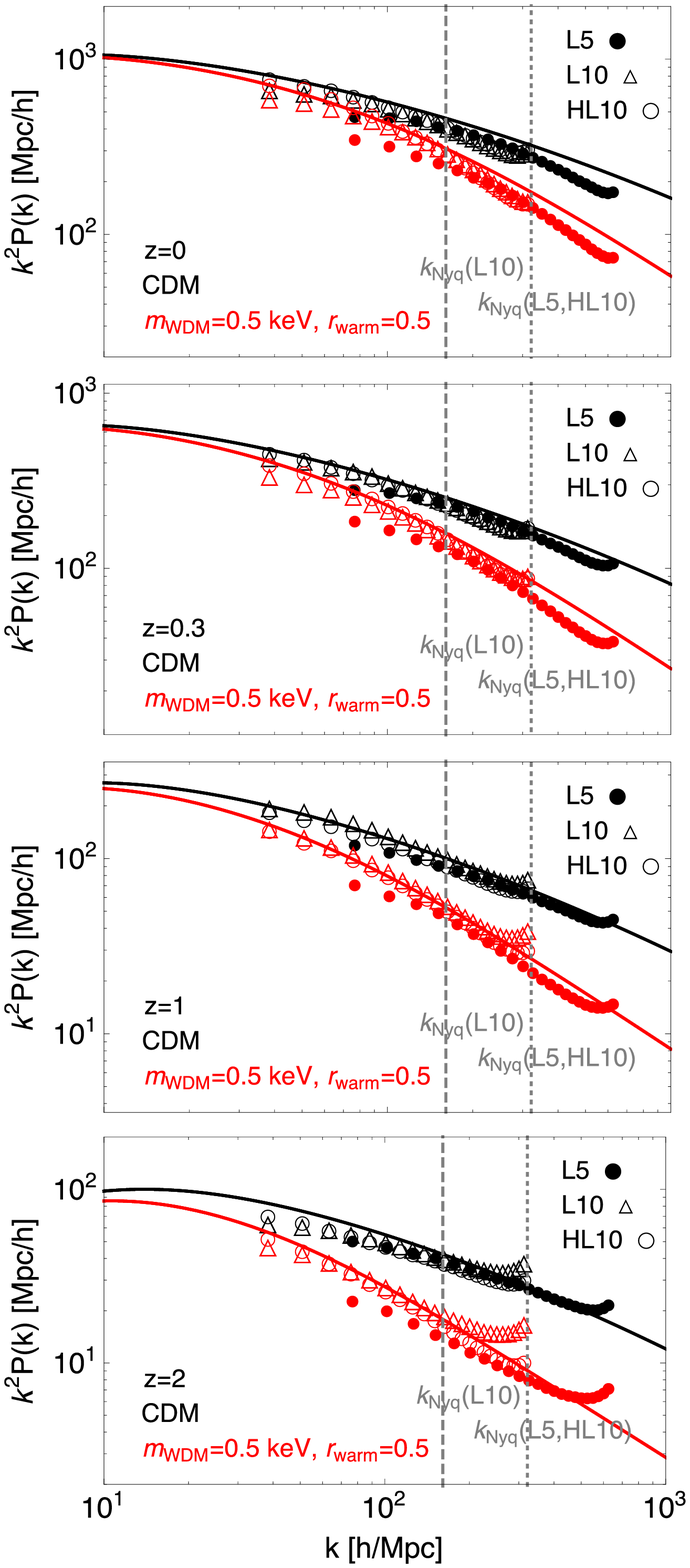}
\caption{Comparison of $P_{\rm MDM}$ (\texttt{halofit} multiplied by $T^{2}$) and $P_{\rm MDM}|_{\rm L5, L10, HL10}$ at $z=0,0.3,1,$ and $2$. We show CDM (black) and MDM(0.5, 0.5) (red) models.
}
\label{fig:nonlinearH}
\EF


\bibliography{mdm_lens}

\begin{thebibliography}{64}%
\makeatletter
\providecommand \@ifxundefined [1]{%
 \@ifx{#1\undefined}
}%
\providecommand \@ifnum [1]{%
 \ifnum #1\expandafter \@firstoftwo
 \else \expandafter \@secondoftwo
 \fi
}%
\providecommand \@ifx [1]{%
 \ifx #1\expandafter \@firstoftwo
 \else \expandafter \@secondoftwo
 \fi
}%
\providecommand \natexlab [1]{#1}%
\providecommand \enquote  [1]{``#1''}%
\providecommand \bibnamefont  [1]{#1}%
\providecommand \bibfnamefont [1]{#1}%
\providecommand \citenamefont [1]{#1}%
\providecommand \href@noop [0]{\@secondoftwo}%
\providecommand \href [0]{\begingroup \@sanitize@url \@href}%
\providecommand \@href[1]{\@@startlink{#1}\@@href}%
\providecommand \@@href[1]{\endgroup#1\@@endlink}%
\providecommand \@sanitize@url [0]{\catcode `\\12\catcode `\$12\catcode
  `\&12\catcode `\#12\catcode `\^12\catcode `\_12\catcode `\%12\relax}%
\providecommand \@@startlink[1]{}%
\providecommand \@@endlink[0]{}%
\providecommand \url  [0]{\begingroup\@sanitize@url \@url }%
\providecommand \@url [1]{\endgroup\@href {#1}{\urlprefix }}%
\providecommand \urlprefix  [0]{URL }%
\providecommand \Eprint [0]{\href }%
\providecommand \doibase [0]{http://dx.doi.org/}%
\providecommand \selectlanguage [0]{\@gobble}%
\providecommand \bibinfo  [0]{\@secondoftwo}%
\providecommand \bibfield  [0]{\@secondoftwo}%
\providecommand \translation [1]{[#1]}%
\providecommand \BibitemOpen [0]{}%
\providecommand \bibitemStop [0]{}%
\providecommand \bibitemNoStop [0]{.\EOS\space}%
\providecommand \EOS [0]{\spacefactor3000\relax}%
\providecommand \BibitemShut  [1]{\csname bibitem#1\endcsname}%
\let\auto@bib@innerbib\@empty
\bibitem [{\citenamefont {Bode}\ \emph {et~al.}(2001)\citenamefont {Bode},
  \citenamefont {Ostriker},\ and\ \citenamefont {Turok}}]{Bode:2000gq}%
  \BibitemOpen
  \bibfield  {author} {\bibinfo {author} {\bibfnamefont {P.}~\bibnamefont
  {Bode}}, \bibinfo {author} {\bibfnamefont {J.~P.}\ \bibnamefont {Ostriker}},
  \ and\ \bibinfo {author} {\bibfnamefont {N.}~\bibnamefont {Turok}},\ }\href
  {\doibase 10.1086/321541} {\bibfield  {journal} {\bibinfo  {journal} {\apj}\
  }\textbf {\bibinfo {volume} {556}},\ \bibinfo {pages} {93} (\bibinfo {year}
  {2001})},\ \Eprint {http://arxiv.org/abs/astro-ph/0010389}
  {arXiv:astro-ph/0010389 [astro-ph]} \BibitemShut {NoStop}%
\bibitem [{\citenamefont {Moore}\ \emph {et~al.}(1999)\citenamefont {Moore},
  \citenamefont {Ghigna}, \citenamefont {Governato}, \citenamefont {Lake},
  \citenamefont {Quinn}, \citenamefont {Stadel},\ and\ \citenamefont
  {Tozzi}}]{Moore:1999nt}%
  \BibitemOpen
  \bibfield  {author} {\bibinfo {author} {\bibfnamefont {B.}~\bibnamefont
  {Moore}}, \bibinfo {author} {\bibfnamefont {S.}~\bibnamefont {Ghigna}},
  \bibinfo {author} {\bibfnamefont {F.}~\bibnamefont {Governato}}, \bibinfo
  {author} {\bibfnamefont {G.}~\bibnamefont {Lake}}, \bibinfo {author}
  {\bibfnamefont {T.~R.}\ \bibnamefont {Quinn}}, \bibinfo {author}
  {\bibfnamefont {J.}~\bibnamefont {Stadel}}, \ and\ \bibinfo {author}
  {\bibfnamefont {P.}~\bibnamefont {Tozzi}},\ }\href {\doibase 10.1086/312287}
  {\bibfield  {journal} {\bibinfo  {journal} {\apj}\ }\textbf {\bibinfo
  {volume} {524}},\ \bibinfo {pages} {L19} (\bibinfo {year} {1999})},\ \Eprint
  {http://arxiv.org/abs/astro-ph/9907411} {arXiv:astro-ph/9907411 [astro-ph]}
  \BibitemShut {NoStop}%
\bibitem [{\citenamefont {Kravtsov}(2010)}]{Kravtsov:2009gi}%
  \BibitemOpen
  \bibfield  {author} {\bibinfo {author} {\bibfnamefont {A.~V.}\ \bibnamefont
  {Kravtsov}},\ }\href {\doibase 10.1155/2010/281913} {\bibfield  {journal}
  {\bibinfo  {journal} {Adv. Astron.}\ }\textbf {\bibinfo {volume} {2010}},\
  \bibinfo {pages} {281913} (\bibinfo {year} {2010})},\ \Eprint
  {http://arxiv.org/abs/0906.3295} {arXiv:0906.3295 [astro-ph.CO]} \BibitemShut
  {NoStop}%
\bibitem [{\citenamefont {Polisensky}\ and\ \citenamefont
  {Ricotti}(2011)}]{Polisensky:2010rw}%
  \BibitemOpen
  \bibfield  {author} {\bibinfo {author} {\bibfnamefont {E.}~\bibnamefont
  {Polisensky}}\ and\ \bibinfo {author} {\bibfnamefont {M.}~\bibnamefont
  {Ricotti}},\ }\href {\doibase 10.1103/PhysRevD.83.043506} {\bibfield
  {journal} {\bibinfo  {journal} {Phys. Rev.}\ }\textbf {\bibinfo {volume}
  {D83}},\ \bibinfo {pages} {043506} (\bibinfo {year} {2011})},\ \Eprint
  {http://arxiv.org/abs/1004.1459} {arXiv:1004.1459 [astro-ph.CO]} \BibitemShut
  {NoStop}%
\bibitem [{\citenamefont {Lovell}\ \emph {et~al.}(2014)\citenamefont {Lovell},
  \citenamefont {Frenk}, \citenamefont {Eke}, \citenamefont {Jenkins},
  \citenamefont {Gao},\ and\ \citenamefont {Theuns}}]{Lovell:2013ola}%
  \BibitemOpen
  \bibfield  {author} {\bibinfo {author} {\bibfnamefont {M.~R.}\ \bibnamefont
  {Lovell}}, \bibinfo {author} {\bibfnamefont {C.~S.}\ \bibnamefont {Frenk}},
  \bibinfo {author} {\bibfnamefont {V.~R.}\ \bibnamefont {Eke}}, \bibinfo
  {author} {\bibfnamefont {A.}~\bibnamefont {Jenkins}}, \bibinfo {author}
  {\bibfnamefont {L.}~\bibnamefont {Gao}}, \ and\ \bibinfo {author}
  {\bibfnamefont {T.}~\bibnamefont {Theuns}},\ }\href {\doibase
  10.1093/mnras/stt2431} {\bibfield  {journal} {\bibinfo  {journal} {Mon. Not.
  Roy. Astron. Soc.}\ }\textbf {\bibinfo {volume} {439}},\ \bibinfo {pages}
  {300} (\bibinfo {year} {2014})},\ \Eprint {http://arxiv.org/abs/1308.1399}
  {arXiv:1308.1399 [astro-ph.CO]} \BibitemShut {NoStop}%
\bibitem [{\citenamefont {Horiuchi}\ \emph {et~al.}(2014)\citenamefont
  {Horiuchi}, \citenamefont {Humphrey}, \citenamefont {Onorbe}, \citenamefont
  {Abazajian}, \citenamefont {Kaplinghat},\ and\ \citenamefont
  {Garrison-Kimmel}}]{Horiuchi:2013noa}%
  \BibitemOpen
  \bibfield  {author} {\bibinfo {author} {\bibfnamefont {S.}~\bibnamefont
  {Horiuchi}}, \bibinfo {author} {\bibfnamefont {P.~J.}\ \bibnamefont
  {Humphrey}}, \bibinfo {author} {\bibfnamefont {J.}~\bibnamefont {Onorbe}},
  \bibinfo {author} {\bibfnamefont {K.~N.}\ \bibnamefont {Abazajian}}, \bibinfo
  {author} {\bibfnamefont {M.}~\bibnamefont {Kaplinghat}}, \ and\ \bibinfo
  {author} {\bibfnamefont {S.}~\bibnamefont {Garrison-Kimmel}},\ }\href
  {\doibase 10.1103/PhysRevD.89.025017} {\bibfield  {journal} {\bibinfo
  {journal} {Phys. Rev.}\ }\textbf {\bibinfo {volume} {D89}},\ \bibinfo {pages}
  {025017} (\bibinfo {year} {2014})},\ \Eprint {http://arxiv.org/abs/1311.0282}
  {arXiv:1311.0282 [astro-ph.CO]} \BibitemShut {NoStop}%
\bibitem [{\citenamefont {Pacucci}\ \emph {et~al.}(2013)\citenamefont
  {Pacucci}, \citenamefont {Mesinger},\ and\ \citenamefont
  {Haiman}}]{Pacucci:2013jfa}%
  \BibitemOpen
  \bibfield  {author} {\bibinfo {author} {\bibfnamefont {F.}~\bibnamefont
  {Pacucci}}, \bibinfo {author} {\bibfnamefont {A.}~\bibnamefont {Mesinger}}, \
  and\ \bibinfo {author} {\bibfnamefont {Z.}~\bibnamefont {Haiman}},\ }\href
  {\doibase 10.1093/mnrasl/slt093} {\bibfield  {journal} {\bibinfo  {journal}
  {Mon. Not. Roy. Astron. Soc.}\ }\textbf {\bibinfo {volume} {435}},\ \bibinfo
  {pages} {L53} (\bibinfo {year} {2013})},\ \Eprint
  {http://arxiv.org/abs/1306.0009} {arXiv:1306.0009 [astro-ph.CO]} \BibitemShut
  {NoStop}%
\bibitem [{\citenamefont {Schultz}\ \emph {et~al.}(2014)\citenamefont
  {Schultz}, \citenamefont {Oñorbe}, \citenamefont {Abazajian},\ and\
  \citenamefont {Bullock}}]{Schultz:2014eia}%
  \BibitemOpen
  \bibfield  {author} {\bibinfo {author} {\bibfnamefont {C.}~\bibnamefont
  {Schultz}}, \bibinfo {author} {\bibfnamefont {J.}~\bibnamefont {Oñorbe}},
  \bibinfo {author} {\bibfnamefont {K.~N.}\ \bibnamefont {Abazajian}}, \ and\
  \bibinfo {author} {\bibfnamefont {J.~S.}\ \bibnamefont {Bullock}},\ }\href
  {\doibase 10.1093/mnras/stu976} {\bibfield  {journal} {\bibinfo  {journal}
  {Mon. Not. Roy. Astron. Soc.}\ }\textbf {\bibinfo {volume} {442}},\ \bibinfo
  {pages} {1597} (\bibinfo {year} {2014})},\ \Eprint
  {http://arxiv.org/abs/1401.3769} {arXiv:1401.3769 [astro-ph.CO]} \BibitemShut
  {NoStop}%
\bibitem [{\citenamefont {Lapi}\ and\ \citenamefont
  {Danese}(2015)}]{Lapi:2015zea}%
  \BibitemOpen
  \bibfield  {author} {\bibinfo {author} {\bibfnamefont {A.}~\bibnamefont
  {Lapi}}\ and\ \bibinfo {author} {\bibfnamefont {L.}~\bibnamefont {Danese}},\
  }\href {\doibase 10.1088/1475-7516/2015/09/003} {\bibfield  {journal}
  {\bibinfo  {journal} {JCAP}\ }\textbf {\bibinfo {volume} {1509}},\ \bibinfo
  {pages} {003} (\bibinfo {year} {2015})},\ \Eprint
  {http://arxiv.org/abs/1508.02147} {arXiv:1508.02147 [astro-ph.CO]}
  \BibitemShut {NoStop}%
\bibitem [{\citenamefont {Menci}\ \emph {et~al.}(2016)\citenamefont {Menci},
  \citenamefont {Sanchez}, \citenamefont {Castellano},\ and\ \citenamefont
  {Grazian}}]{Menci:2016eww}%
  \BibitemOpen
  \bibfield  {author} {\bibinfo {author} {\bibfnamefont {N.}~\bibnamefont
  {Menci}}, \bibinfo {author} {\bibfnamefont {N.~G.}\ \bibnamefont {Sanchez}},
  \bibinfo {author} {\bibfnamefont {M.}~\bibnamefont {Castellano}}, \ and\
  \bibinfo {author} {\bibfnamefont {A.}~\bibnamefont {Grazian}},\ }\href
  {\doibase 10.3847/0004-637X/818/1/90} {\bibfield  {journal} {\bibinfo
  {journal} {Astrophys. J.}\ }\textbf {\bibinfo {volume} {818}},\ \bibinfo
  {pages} {90} (\bibinfo {year} {2016})},\ \Eprint
  {http://arxiv.org/abs/1601.01820} {arXiv:1601.01820 [astro-ph.CO]}
  \BibitemShut {NoStop}%
\bibitem [{\citenamefont {de~Souza}\ \emph {et~al.}(2013)\citenamefont
  {de~Souza}, \citenamefont {Mesinger}, \citenamefont {Ferrara}, \citenamefont
  {Haiman}, \citenamefont {Perna},\ and\ \citenamefont
  {Yoshida}}]{deSouza:2013hsj}%
  \BibitemOpen
  \bibfield  {author} {\bibinfo {author} {\bibfnamefont {R.~S.}\ \bibnamefont
  {de~Souza}}, \bibinfo {author} {\bibfnamefont {A.}~\bibnamefont {Mesinger}},
  \bibinfo {author} {\bibfnamefont {A.}~\bibnamefont {Ferrara}}, \bibinfo
  {author} {\bibfnamefont {Z.}~\bibnamefont {Haiman}}, \bibinfo {author}
  {\bibfnamefont {R.}~\bibnamefont {Perna}}, \ and\ \bibinfo {author}
  {\bibfnamefont {N.}~\bibnamefont {Yoshida}},\ }\href {\doibase
  10.1093/mnras/stt674} {\bibfield  {journal} {\bibinfo  {journal} {Mon. Not.
  Roy. Astron. Soc.}\ }\textbf {\bibinfo {volume} {432}},\ \bibinfo {pages}
  {3218} (\bibinfo {year} {2013})},\ \Eprint {http://arxiv.org/abs/1303.5060}
  {arXiv:1303.5060 [astro-ph.CO]} \BibitemShut {NoStop}%
\bibitem [{\citenamefont {Mesinger}\ \emph {et~al.}(2005)\citenamefont
  {Mesinger}, \citenamefont {Perna},\ and\ \citenamefont
  {Haiman}}]{Mesinger:2005ah}%
  \BibitemOpen
  \bibfield  {author} {\bibinfo {author} {\bibfnamefont {A.}~\bibnamefont
  {Mesinger}}, \bibinfo {author} {\bibfnamefont {R.}~\bibnamefont {Perna}}, \
  and\ \bibinfo {author} {\bibfnamefont {Z.}~\bibnamefont {Haiman}},\ }\href
  {\doibase 10.1086/428770} {\bibfield  {journal} {\bibinfo  {journal}
  {Astrophys. J.}\ }\textbf {\bibinfo {volume} {623}},\ \bibinfo {pages} {1}
  (\bibinfo {year} {2005})},\ \Eprint {http://arxiv.org/abs/astro-ph/0501233}
  {arXiv:astro-ph/0501233 [astro-ph]} \BibitemShut {NoStop}%
\bibitem [{\citenamefont {Pandolfi}\ \emph {et~al.}(2014)\citenamefont
  {Pandolfi}, \citenamefont {Evoli}, \citenamefont {Ferrara},\ and\
  \citenamefont {Villaescusa-Navarro}}]{Pandolfi:2014rea}%
  \BibitemOpen
  \bibfield  {author} {\bibinfo {author} {\bibfnamefont {S.}~\bibnamefont
  {Pandolfi}}, \bibinfo {author} {\bibfnamefont {C.}~\bibnamefont {Evoli}},
  \bibinfo {author} {\bibfnamefont {A.}~\bibnamefont {Ferrara}}, \ and\
  \bibinfo {author} {\bibfnamefont {F.}~\bibnamefont {Villaescusa-Navarro}},\
  }\href {\doibase 10.1093/mnras/stu785} {\bibfield  {journal} {\bibinfo
  {journal} {Mon. Not. Roy. Astron. Soc.}\ }\textbf {\bibinfo {volume} {442}},\
  \bibinfo {pages} {13} (\bibinfo {year} {2014})},\ \Eprint
  {http://arxiv.org/abs/1403.2185} {arXiv:1403.2185 [astro-ph.CO]} \BibitemShut
  {NoStop}%
\bibitem [{\citenamefont {Barkana}\ \emph {et~al.}(2001)\citenamefont
  {Barkana}, \citenamefont {Haiman},\ and\ \citenamefont
  {Ostriker}}]{Barkana:2001gr}%
  \BibitemOpen
  \bibfield  {author} {\bibinfo {author} {\bibfnamefont {R.}~\bibnamefont
  {Barkana}}, \bibinfo {author} {\bibfnamefont {Z.}~\bibnamefont {Haiman}}, \
  and\ \bibinfo {author} {\bibfnamefont {J.~P.}\ \bibnamefont {Ostriker}},\
  }\href {\doibase 10.1086/322393} {\bibfield  {journal} {\bibinfo  {journal}
  {Astrophys. J.}\ }\textbf {\bibinfo {volume} {558}},\ \bibinfo {pages} {482}
  (\bibinfo {year} {2001})},\ \Eprint {http://arxiv.org/abs/astro-ph/0102304}
  {arXiv:astro-ph/0102304 [astro-ph]} \BibitemShut {NoStop}%
\bibitem [{\citenamefont {Viel}\ \emph {et~al.}(2005)\citenamefont {Viel},
  \citenamefont {Lesgourgues}, \citenamefont {Haehnelt}, \citenamefont
  {Matarrese},\ and\ \citenamefont {Riotto}}]{Viel:2005qj}%
  \BibitemOpen
  \bibfield  {author} {\bibinfo {author} {\bibfnamefont {M.}~\bibnamefont
  {Viel}}, \bibinfo {author} {\bibfnamefont {J.}~\bibnamefont {Lesgourgues}},
  \bibinfo {author} {\bibfnamefont {M.~G.}\ \bibnamefont {Haehnelt}}, \bibinfo
  {author} {\bibfnamefont {S.}~\bibnamefont {Matarrese}}, \ and\ \bibinfo
  {author} {\bibfnamefont {A.}~\bibnamefont {Riotto}},\ }\href {\doibase
  10.1103/PhysRevD.71.063534} {\bibfield  {journal} {\bibinfo  {journal} {Phys.
  Rev.}\ }\textbf {\bibinfo {volume} {D71}},\ \bibinfo {pages} {063534}
  (\bibinfo {year} {2005})},\ \Eprint {http://arxiv.org/abs/astro-ph/0501562}
  {arXiv:astro-ph/0501562 [astro-ph]} \BibitemShut {NoStop}%
\bibitem [{\citenamefont {Seljak}\ \emph {et~al.}(2006)\citenamefont {Seljak},
  \citenamefont {Makarov}, \citenamefont {McDonald},\ and\ \citenamefont
  {Trac}}]{Seljak:2006qw}%
  \BibitemOpen
  \bibfield  {author} {\bibinfo {author} {\bibfnamefont {U.}~\bibnamefont
  {Seljak}}, \bibinfo {author} {\bibfnamefont {A.}~\bibnamefont {Makarov}},
  \bibinfo {author} {\bibfnamefont {P.}~\bibnamefont {McDonald}}, \ and\
  \bibinfo {author} {\bibfnamefont {H.}~\bibnamefont {Trac}},\ }\href {\doibase
  10.1103/PhysRevLett.97.191303} {\bibfield  {journal} {\bibinfo  {journal}
  {Phys. Rev. Lett.}\ }\textbf {\bibinfo {volume} {97}},\ \bibinfo {pages}
  {191303} (\bibinfo {year} {2006})},\ \Eprint
  {http://arxiv.org/abs/astro-ph/0602430} {arXiv:astro-ph/0602430 [astro-ph]}
  \BibitemShut {NoStop}%
\bibitem [{\citenamefont {Viel}\ \emph {et~al.}(2006)\citenamefont {Viel},
  \citenamefont {Lesgourgues}, \citenamefont {Haehnelt}, \citenamefont
  {Matarrese},\ and\ \citenamefont {Riotto}}]{Viel:2006kd}%
  \BibitemOpen
  \bibfield  {author} {\bibinfo {author} {\bibfnamefont {M.}~\bibnamefont
  {Viel}}, \bibinfo {author} {\bibfnamefont {J.}~\bibnamefont {Lesgourgues}},
  \bibinfo {author} {\bibfnamefont {M.~G.}\ \bibnamefont {Haehnelt}}, \bibinfo
  {author} {\bibfnamefont {S.}~\bibnamefont {Matarrese}}, \ and\ \bibinfo
  {author} {\bibfnamefont {A.}~\bibnamefont {Riotto}},\ }\href {\doibase
  10.1103/PhysRevLett.97.071301} {\bibfield  {journal} {\bibinfo  {journal}
  {Phys. Rev. Lett.}\ }\textbf {\bibinfo {volume} {97}},\ \bibinfo {pages}
  {071301} (\bibinfo {year} {2006})},\ \Eprint
  {http://arxiv.org/abs/astro-ph/0605706} {arXiv:astro-ph/0605706 [astro-ph]}
  \BibitemShut {NoStop}%
\bibitem [{\citenamefont {Boyarsky}\ \emph {et~al.}(2009)\citenamefont
  {Boyarsky}, \citenamefont {Lesgourgues}, \citenamefont {Ruchayskiy},\ and\
  \citenamefont {Viel}}]{Boyarsky:2008xj}%
  \BibitemOpen
  \bibfield  {author} {\bibinfo {author} {\bibfnamefont {A.}~\bibnamefont
  {Boyarsky}}, \bibinfo {author} {\bibfnamefont {J.}~\bibnamefont
  {Lesgourgues}}, \bibinfo {author} {\bibfnamefont {O.}~\bibnamefont
  {Ruchayskiy}}, \ and\ \bibinfo {author} {\bibfnamefont {M.}~\bibnamefont
  {Viel}},\ }\href {\doibase 10.1088/1475-7516/2009/05/012} {\bibfield
  {journal} {\bibinfo  {journal} {JCAP}\ }\textbf {\bibinfo {volume} {0905}},\
  \bibinfo {pages} {012} (\bibinfo {year} {2009})},\ \Eprint
  {http://arxiv.org/abs/0812.0010} {arXiv:0812.0010 [astro-ph]} \BibitemShut
  {NoStop}%
\bibitem [{\citenamefont {Viel}\ \emph {et~al.}(2013)\citenamefont {Viel},
  \citenamefont {Becker}, \citenamefont {Bolton},\ and\ \citenamefont
  {Haehnelt}}]{Viel:2013apy}%
  \BibitemOpen
  \bibfield  {author} {\bibinfo {author} {\bibfnamefont {M.}~\bibnamefont
  {Viel}}, \bibinfo {author} {\bibfnamefont {G.~D.}\ \bibnamefont {Becker}},
  \bibinfo {author} {\bibfnamefont {J.~S.}\ \bibnamefont {Bolton}}, \ and\
  \bibinfo {author} {\bibfnamefont {M.~G.}\ \bibnamefont {Haehnelt}},\ }\href
  {\doibase 10.1103/PhysRevD.88.043502} {\bibfield  {journal} {\bibinfo
  {journal} {Phys. Rev.}\ }\textbf {\bibinfo {volume} {D88}},\ \bibinfo {pages}
  {043502} (\bibinfo {year} {2013})},\ \Eprint {http://arxiv.org/abs/1306.2314}
  {arXiv:1306.2314 [astro-ph.CO]} \BibitemShut {NoStop}%
\bibitem [{\citenamefont {Schneider}\ \emph {et~al.}(2014)\citenamefont
  {Schneider}, \citenamefont {Anderhalden}, \citenamefont {Maccio},\ and\
  \citenamefont {Diemand}}]{Schneider:2013wwa}%
  \BibitemOpen
  \bibfield  {author} {\bibinfo {author} {\bibfnamefont {A.}~\bibnamefont
  {Schneider}}, \bibinfo {author} {\bibfnamefont {D.}~\bibnamefont
  {Anderhalden}}, \bibinfo {author} {\bibfnamefont {A.}~\bibnamefont {Maccio}},
  \ and\ \bibinfo {author} {\bibfnamefont {J.}~\bibnamefont {Diemand}},\ }\href
  {\doibase 10.1093/mnrasl/slu034} {\bibfield  {journal} {\bibinfo  {journal}
  {\mnras}\ }\textbf {\bibinfo {volume} {441}},\ \bibinfo {pages} {6} (\bibinfo
  {year} {2014})},\ \Eprint {http://arxiv.org/abs/1309.5960} {arXiv:1309.5960
  [astro-ph.CO]} \BibitemShut {NoStop}%
\bibitem [{\citenamefont {Anderhalden}\ \emph {et~al.}(2012)\citenamefont
  {Anderhalden}, \citenamefont {Diemand}, \citenamefont {Bertone},
  \citenamefont {Maccio},\ and\ \citenamefont
  {Schneider}}]{Anderhalden:2012qt}%
  \BibitemOpen
  \bibfield  {author} {\bibinfo {author} {\bibfnamefont {D.}~\bibnamefont
  {Anderhalden}}, \bibinfo {author} {\bibfnamefont {J.}~\bibnamefont
  {Diemand}}, \bibinfo {author} {\bibfnamefont {G.}~\bibnamefont {Bertone}},
  \bibinfo {author} {\bibfnamefont {A.~V.}\ \bibnamefont {Maccio}}, \ and\
  \bibinfo {author} {\bibfnamefont {A.}~\bibnamefont {Schneider}},\ }\href
  {\doibase 10.1088/1475-7516/2012/10/047} {\bibfield  {journal} {\bibinfo
  {journal} {JCAP}\ }\textbf {\bibinfo {volume} {1210}},\ \bibinfo {pages}
  {047} (\bibinfo {year} {2012})},\ \Eprint {http://arxiv.org/abs/1206.3788}
  {arXiv:1206.3788 [astro-ph.CO]} \BibitemShut {NoStop}%
\bibitem [{\citenamefont {Anderhalden}\ \emph {et~al.}(2013)\citenamefont
  {Anderhalden}, \citenamefont {Schneider}, \citenamefont {Maccio},
  \citenamefont {Diemand},\ and\ \citenamefont {Bertone}}]{Anderhalden:2012jc}%
  \BibitemOpen
  \bibfield  {author} {\bibinfo {author} {\bibfnamefont {D.}~\bibnamefont
  {Anderhalden}}, \bibinfo {author} {\bibfnamefont {A.}~\bibnamefont
  {Schneider}}, \bibinfo {author} {\bibfnamefont {A.~V.}\ \bibnamefont
  {Maccio}}, \bibinfo {author} {\bibfnamefont {J.}~\bibnamefont {Diemand}}, \
  and\ \bibinfo {author} {\bibfnamefont {G.}~\bibnamefont {Bertone}},\ }\href
  {\doibase 10.1088/1475-7516/2013/03/014} {\bibfield  {journal} {\bibinfo
  {journal} {JCAP}\ }\textbf {\bibinfo {volume} {1303}},\ \bibinfo {pages}
  {014} (\bibinfo {year} {2013})},\ \Eprint {http://arxiv.org/abs/1212.2967}
  {arXiv:1212.2967 [astro-ph.CO]} \BibitemShut {NoStop}%
\bibitem [{\citenamefont {Bulbul}\ \emph {et~al.}(2014)\citenamefont {Bulbul},
  \citenamefont {Markevitch}, \citenamefont {Foster}, \citenamefont {Smith},
  \citenamefont {Loewenstein},\ and\ \citenamefont {Randall}}]{Bulbul:2014sua}%
  \BibitemOpen
  \bibfield  {author} {\bibinfo {author} {\bibfnamefont {E.}~\bibnamefont
  {Bulbul}}, \bibinfo {author} {\bibfnamefont {M.}~\bibnamefont {Markevitch}},
  \bibinfo {author} {\bibfnamefont {A.}~\bibnamefont {Foster}}, \bibinfo
  {author} {\bibfnamefont {R.~K.}\ \bibnamefont {Smith}}, \bibinfo {author}
  {\bibfnamefont {M.}~\bibnamefont {Loewenstein}}, \ and\ \bibinfo {author}
  {\bibfnamefont {S.~W.}\ \bibnamefont {Randall}},\ }\href {\doibase
  10.1088/0004-637X/789/1/13} {\bibfield  {journal} {\bibinfo  {journal}
  {Astrophys. J.}\ }\textbf {\bibinfo {volume} {789}},\ \bibinfo {pages} {13}
  (\bibinfo {year} {2014})},\ \Eprint {http://arxiv.org/abs/1402.2301}
  {arXiv:1402.2301 [astro-ph.CO]} \BibitemShut {NoStop}%
\bibitem [{\citenamefont {Boyarsky}\ \emph {et~al.}(2014)\citenamefont
  {Boyarsky}, \citenamefont {Ruchayskiy}, \citenamefont {Iakubovskyi},\ and\
  \citenamefont {Franse}}]{Boyarsky:2014jta}%
  \BibitemOpen
  \bibfield  {author} {\bibinfo {author} {\bibfnamefont {A.}~\bibnamefont
  {Boyarsky}}, \bibinfo {author} {\bibfnamefont {O.}~\bibnamefont
  {Ruchayskiy}}, \bibinfo {author} {\bibfnamefont {D.}~\bibnamefont
  {Iakubovskyi}}, \ and\ \bibinfo {author} {\bibfnamefont {J.}~\bibnamefont
  {Franse}},\ }\href {\doibase 10.1103/PhysRevLett.113.251301} {\bibfield
  {journal} {\bibinfo  {journal} {Phys. Rev. Lett.}\ }\textbf {\bibinfo
  {volume} {113}},\ \bibinfo {pages} {251301} (\bibinfo {year} {2014})},\
  \Eprint {http://arxiv.org/abs/1402.4119} {arXiv:1402.4119 [astro-ph.CO]}
  \BibitemShut {NoStop}%
\bibitem [{\citenamefont {Tamura}\ \emph {et~al.}(2015)\citenamefont {Tamura},
  \citenamefont {Iizuka}, \citenamefont {Maeda}, \citenamefont {Mitsuda},\ and\
  \citenamefont {Yamasaki}}]{Tamura:2014mta}%
  \BibitemOpen
  \bibfield  {author} {\bibinfo {author} {\bibfnamefont {T.}~\bibnamefont
  {Tamura}}, \bibinfo {author} {\bibfnamefont {R.}~\bibnamefont {Iizuka}},
  \bibinfo {author} {\bibfnamefont {Y.}~\bibnamefont {Maeda}}, \bibinfo
  {author} {\bibfnamefont {K.}~\bibnamefont {Mitsuda}}, \ and\ \bibinfo
  {author} {\bibfnamefont {N.~Y.}\ \bibnamefont {Yamasaki}},\ }\href {\doibase
  10.1093/pasj/psu156} {\bibfield  {journal} {\bibinfo  {journal} {Publ.
  Astron. Soc. Jap.}\ }\textbf {\bibinfo {volume} {67}},\ \bibinfo {pages} {23}
  (\bibinfo {year} {2015})},\ \Eprint {http://arxiv.org/abs/1412.1869}
  {arXiv:1412.1869 [astro-ph.HE]} \BibitemShut {NoStop}%
\bibitem [{\citenamefont {Kusenko}(2009)}]{Kusenko:2009up}%
  \BibitemOpen
  \bibfield  {author} {\bibinfo {author} {\bibfnamefont {A.}~\bibnamefont
  {Kusenko}},\ }\href {\doibase 10.1016/j.physrep.2009.07.004} {\bibfield
  {journal} {\bibinfo  {journal} {Phys. Rept.}\ }\textbf {\bibinfo {volume}
  {481}},\ \bibinfo {pages} {1} (\bibinfo {year} {2009})},\ \Eprint
  {http://arxiv.org/abs/0906.2968} {arXiv:0906.2968 [hep-ph]} \BibitemShut
  {NoStop}%
\bibitem [{\citenamefont {Harada}\ and\ \citenamefont
  {Kamada}(2016)}]{Harada:2014lma}%
  \BibitemOpen
  \bibfield  {author} {\bibinfo {author} {\bibfnamefont {A.}~\bibnamefont
  {Harada}}\ and\ \bibinfo {author} {\bibfnamefont {A.}~\bibnamefont
  {Kamada}},\ }\href {\doibase 10.1088/1475-7516/2016/01/031} {\bibfield
  {journal} {\bibinfo  {journal} {JCAP}\ }\textbf {\bibinfo {volume} {1601}},\
  \bibinfo {pages} {031} (\bibinfo {year} {2016})},\ \Eprint
  {http://arxiv.org/abs/1412.1592} {arXiv:1412.1592 [astro-ph.CO]} \BibitemShut
  {NoStop}%
\bibitem [{\citenamefont {Dodelson}\ and\ \citenamefont
  {Widrow}(1994)}]{Dodelson:1993je}%
  \BibitemOpen
  \bibfield  {author} {\bibinfo {author} {\bibfnamefont {S.}~\bibnamefont
  {Dodelson}}\ and\ \bibinfo {author} {\bibfnamefont {L.~M.}\ \bibnamefont
  {Widrow}},\ }\href {\doibase 10.1103/PhysRevLett.72.17} {\bibfield  {journal}
  {\bibinfo  {journal} {Phys. Rev. Lett.}\ }\textbf {\bibinfo {volume} {72}},\
  \bibinfo {pages} {17} (\bibinfo {year} {1994})},\ \Eprint
  {http://arxiv.org/abs/hep-ph/9303287} {arXiv:hep-ph/9303287 [hep-ph]}
  \BibitemShut {NoStop}%
\bibitem [{\citenamefont {{Mao}}\ and\ \citenamefont
  {{Schneider}}(1998)}]{mao1998}%
  \BibitemOpen
  \bibfield  {author} {\bibinfo {author} {\bibfnamefont {S.}~\bibnamefont
  {{Mao}}}\ and\ \bibinfo {author} {\bibfnamefont {P.}~\bibnamefont
  {{Schneider}}},\ }\href {\doibase 10.1046/j.1365-8711.1998.01319.x}
  {\bibfield  {journal} {\bibinfo  {journal} {\mnras}\ }\textbf {\bibinfo
  {volume} {295}},\ \bibinfo {pages} {587} (\bibinfo {year} {1998})},\ \Eprint
  {http://arxiv.org/abs/astro-ph/9707187} {astro-ph/9707187} \BibitemShut
  {NoStop}%
\bibitem [{\citenamefont {{Metcalf}}\ and\ \citenamefont
  {{Madau}}(2001)}]{metcalf2001}%
  \BibitemOpen
  \bibfield  {author} {\bibinfo {author} {\bibfnamefont {R.~B.}\ \bibnamefont
  {{Metcalf}}}\ and\ \bibinfo {author} {\bibfnamefont {P.}~\bibnamefont
  {{Madau}}},\ }\href {\doibase 10.1086/323695} {\bibfield  {journal} {\bibinfo
   {journal} {\apj}\ }\textbf {\bibinfo {volume} {563}},\ \bibinfo {pages} {9}
  (\bibinfo {year} {2001})},\ \Eprint {http://arxiv.org/abs/astro-ph/0108224}
  {astro-ph/0108224} \BibitemShut {NoStop}%
\bibitem [{\citenamefont {{Chiba}}(2002)}]{chiba2002}%
  \BibitemOpen
  \bibfield  {author} {\bibinfo {author} {\bibfnamefont {M.}~\bibnamefont
  {{Chiba}}},\ }\href {\doibase 10.1086/324493} {\bibfield  {journal} {\bibinfo
   {journal} {\apj}\ }\textbf {\bibinfo {volume} {565}},\ \bibinfo {pages} {17}
  (\bibinfo {year} {2002})},\ \Eprint {http://arxiv.org/abs/astro-ph/0109499}
  {astro-ph/0109499} \BibitemShut {NoStop}%
\bibitem [{\citenamefont {{Dalal}}\ and\ \citenamefont
  {{Kochanek}}(2002)}]{dalal-kochanek2002}%
  \BibitemOpen
  \bibfield  {author} {\bibinfo {author} {\bibfnamefont {N.}~\bibnamefont
  {{Dalal}}}\ and\ \bibinfo {author} {\bibfnamefont {C.~S.}\ \bibnamefont
  {{Kochanek}}},\ }\href {\doibase 10.1086/340303} {\bibfield  {journal}
  {\bibinfo  {journal} {\apj}\ }\textbf {\bibinfo {volume} {572}},\ \bibinfo
  {pages} {25} (\bibinfo {year} {2002})},\ \Eprint
  {http://arxiv.org/abs/astro-ph/0111456} {astro-ph/0111456} \BibitemShut
  {NoStop}%
\bibitem [{\citenamefont {Keeton}\ \emph {et~al.}(2003)\citenamefont {Keeton},
  \citenamefont {Gaudi},\ and\ \citenamefont {Petters}}]{keeton2003}%
  \BibitemOpen
  \bibfield  {author} {\bibinfo {author} {\bibfnamefont {C.~R.}\ \bibnamefont
  {Keeton}}, \bibinfo {author} {\bibfnamefont {B.~S.}\ \bibnamefont {Gaudi}}, \
  and\ \bibinfo {author} {\bibfnamefont {A.~O.}\ \bibnamefont {Petters}},\
  }\href {\doibase 10.1086/378934} {\bibfield  {journal} {\bibinfo  {journal}
  {\apj}\ }\textbf {\bibinfo {volume} {598}},\ \bibinfo {pages} {138} (\bibinfo
  {year} {2003})},\ \bibinfo {note} {part 1 745KW Times Cited:65 Cited
  References Count:121}\BibitemShut {NoStop}%
\bibitem [{\citenamefont {{Inoue}}\ and\ \citenamefont
  {{Chiba}}(2003)}]{inoue-chiba2003}%
  \BibitemOpen
  \bibfield  {author} {\bibinfo {author} {\bibfnamefont {K.~T.}\ \bibnamefont
  {{Inoue}}}\ and\ \bibinfo {author} {\bibfnamefont {M.}~\bibnamefont
  {{Chiba}}},\ }\href {\doibase 10.1086/377247} {\bibfield  {journal} {\bibinfo
   {journal} {\apj}\ }\textbf {\bibinfo {volume} {591}},\ \bibinfo {pages}
  {L83} (\bibinfo {year} {2003})},\ \Eprint
  {http://arxiv.org/abs/astro-ph/0304474} {astro-ph/0304474} \BibitemShut
  {NoStop}%
\bibitem [{\citenamefont {Kochanek}\ and\ \citenamefont
  {Dalal}(2004)}]{kochanek2004}%
  \BibitemOpen
  \bibfield  {author} {\bibinfo {author} {\bibfnamefont {C.~S.}\ \bibnamefont
  {Kochanek}}\ and\ \bibinfo {author} {\bibfnamefont {N.}~\bibnamefont
  {Dalal}},\ }\href {\doibase 10.1086/421436} {\bibfield  {journal} {\bibinfo
  {journal} {\apj}\ }\textbf {\bibinfo {volume} {610}},\ \bibinfo {pages} {69}
  (\bibinfo {year} {2004})}\BibitemShut {NoStop}%
\bibitem [{\citenamefont {{Metcalf}}\ \emph {et~al.}(2004)\citenamefont
  {{Metcalf}}, \citenamefont {{Moustakas}}, \citenamefont {{Bunker}},\ and\
  \citenamefont {{Parry}}}]{metcalf2004}%
  \BibitemOpen
  \bibfield  {author} {\bibinfo {author} {\bibfnamefont {R.~B.}\ \bibnamefont
  {{Metcalf}}}, \bibinfo {author} {\bibfnamefont {L.~A.}\ \bibnamefont
  {{Moustakas}}}, \bibinfo {author} {\bibfnamefont {A.~J.}\ \bibnamefont
  {{Bunker}}}, \ and\ \bibinfo {author} {\bibfnamefont {I.~R.}\ \bibnamefont
  {{Parry}}},\ }\href {\doibase 10.1086/383243} {\bibfield  {journal} {\bibinfo
   {journal} {\apj}\ }\textbf {\bibinfo {volume} {607}},\ \bibinfo {pages} {43}
  (\bibinfo {year} {2004})},\ \Eprint {http://arxiv.org/abs/astro-ph/0309738}
  {astro-ph/0309738} \BibitemShut {NoStop}%
\bibitem [{\citenamefont {Chiba}\ \emph {et~al.}(2005)\citenamefont {Chiba},
  \citenamefont {Minezaki}, \citenamefont {Kashikawa}, \citenamefont {Kataza},\
  and\ \citenamefont {Inoue}}]{chiba2005}%
  \BibitemOpen
  \bibfield  {author} {\bibinfo {author} {\bibfnamefont {M.}~\bibnamefont
  {Chiba}}, \bibinfo {author} {\bibfnamefont {T.}~\bibnamefont {Minezaki}},
  \bibinfo {author} {\bibfnamefont {N.}~\bibnamefont {Kashikawa}}, \bibinfo
  {author} {\bibfnamefont {H.}~\bibnamefont {Kataza}}, \ and\ \bibinfo {author}
  {\bibfnamefont {K.~T.}\ \bibnamefont {Inoue}},\ }\href {\doibase
  10.1086/430403} {\bibfield  {journal} {\bibinfo  {journal} {\apj}\ }\textbf
  {\bibinfo {volume} {627}},\ \bibinfo {pages} {53} (\bibinfo {year} {2005})},\
  \bibinfo {note} {part 1 939PZ Times Cited:41 Cited References
  Count:65}\BibitemShut {NoStop}%
\bibitem [{\citenamefont {{Sugai}}\ \emph {et~al.}(2007)\citenamefont
  {{Sugai}}, \citenamefont {{Kawai}}, \citenamefont {{Shimono}}, \citenamefont
  {{Hattori}}, \citenamefont {{Kosugi}}, \citenamefont {{Kashikawa}},
  \citenamefont {{Inoue}},\ and\ \citenamefont {{Chiba}}}]{sugai2007}%
  \BibitemOpen
  \bibfield  {author} {\bibinfo {author} {\bibfnamefont {H.}~\bibnamefont
  {{Sugai}}}, \bibinfo {author} {\bibfnamefont {A.}~\bibnamefont {{Kawai}}},
  \bibinfo {author} {\bibfnamefont {A.}~\bibnamefont {{Shimono}}}, \bibinfo
  {author} {\bibfnamefont {T.}~\bibnamefont {{Hattori}}}, \bibinfo {author}
  {\bibfnamefont {G.}~\bibnamefont {{Kosugi}}}, \bibinfo {author}
  {\bibfnamefont {N.}~\bibnamefont {{Kashikawa}}}, \bibinfo {author}
  {\bibfnamefont {K.~T.}\ \bibnamefont {{Inoue}}}, \ and\ \bibinfo {author}
  {\bibfnamefont {M.}~\bibnamefont {{Chiba}}},\ }\href {\doibase
  10.1086/513731} {\bibfield  {journal} {\bibinfo  {journal} {\apj}\ }\textbf
  {\bibinfo {volume} {660}},\ \bibinfo {pages} {1016} (\bibinfo {year}
  {2007})},\ \Eprint {http://arxiv.org/abs/astro-ph/0702392} {astro-ph/0702392}
  \BibitemShut {NoStop}%
\bibitem [{\citenamefont {{McKean}}\ \emph {et~al.}(2007)\citenamefont
  {{McKean}}, \citenamefont {{Koopmans}}, \citenamefont {{Flack}},
  \citenamefont {{Fassnacht}}, \citenamefont {{Thompson}}, \citenamefont
  {{Matthews}}, \citenamefont {{Blandford}}, \citenamefont {{Readhead}},\ and\
  \citenamefont {{Soifer}}}]{mckean2007}%
  \BibitemOpen
  \bibfield  {author} {\bibinfo {author} {\bibfnamefont {J.~P.}\ \bibnamefont
  {{McKean}}}, \bibinfo {author} {\bibfnamefont {L.~V.~E.}\ \bibnamefont
  {{Koopmans}}}, \bibinfo {author} {\bibfnamefont {C.~E.}\ \bibnamefont
  {{Flack}}}, \bibinfo {author} {\bibfnamefont {C.~D.}\ \bibnamefont
  {{Fassnacht}}}, \bibinfo {author} {\bibfnamefont {D.}~\bibnamefont
  {{Thompson}}}, \bibinfo {author} {\bibfnamefont {K.}~\bibnamefont
  {{Matthews}}}, \bibinfo {author} {\bibfnamefont {R.~D.}\ \bibnamefont
  {{Blandford}}}, \bibinfo {author} {\bibfnamefont {A.~C.~S.}\ \bibnamefont
  {{Readhead}}}, \ and\ \bibinfo {author} {\bibfnamefont {B.~T.}\ \bibnamefont
  {{Soifer}}},\ }\href {\doibase 10.1111/j.1365-2966.2007.11744.x} {\bibfield
  {journal} {\bibinfo  {journal} {\mnras}\ }\textbf {\bibinfo {volume} {378}},\
  \bibinfo {pages} {109} (\bibinfo {year} {2007})},\ \Eprint
  {http://arxiv.org/abs/astro-ph/0611215} {astro-ph/0611215} \BibitemShut
  {NoStop}%
\bibitem [{\citenamefont {{More}}\ \emph {et~al.}(2009)\citenamefont {{More}},
  \citenamefont {{McKean}}, \citenamefont {{More}}, \citenamefont {{Porcas}},
  \citenamefont {{Koopmans}},\ and\ \citenamefont {{Garrett}}}]{more2009}%
  \BibitemOpen
  \bibfield  {author} {\bibinfo {author} {\bibfnamefont {A.}~\bibnamefont
  {{More}}}, \bibinfo {author} {\bibfnamefont {J.~P.}\ \bibnamefont
  {{McKean}}}, \bibinfo {author} {\bibfnamefont {S.}~\bibnamefont {{More}}},
  \bibinfo {author} {\bibfnamefont {R.~W.}\ \bibnamefont {{Porcas}}}, \bibinfo
  {author} {\bibfnamefont {L.~V.~E.}\ \bibnamefont {{Koopmans}}}, \ and\
  \bibinfo {author} {\bibfnamefont {M.~A.}\ \bibnamefont {{Garrett}}},\ }\href
  {\doibase 10.1111/j.1365-2966.2008.14342.x} {\bibfield  {journal} {\bibinfo
  {journal} {\mnras}\ }\textbf {\bibinfo {volume} {394}},\ \bibinfo {pages}
  {174} (\bibinfo {year} {2009})},\ \Eprint {http://arxiv.org/abs/0810.5341}
  {arXiv:0810.5341} \BibitemShut {NoStop}%
\bibitem [{\citenamefont {{Minezaki}}\ \emph {et~al.}(2009)\citenamefont
  {{Minezaki}}, \citenamefont {{Chiba}}, \citenamefont {{Kashikawa}},
  \citenamefont {{Inoue}},\ and\ \citenamefont {{Kataza}}}]{minezaki2009}%
  \BibitemOpen
  \bibfield  {author} {\bibinfo {author} {\bibfnamefont {T.}~\bibnamefont
  {{Minezaki}}}, \bibinfo {author} {\bibfnamefont {M.}~\bibnamefont {{Chiba}}},
  \bibinfo {author} {\bibfnamefont {N.}~\bibnamefont {{Kashikawa}}}, \bibinfo
  {author} {\bibfnamefont {K.~T.}\ \bibnamefont {{Inoue}}}, \ and\ \bibinfo
  {author} {\bibfnamefont {H.}~\bibnamefont {{Kataza}}},\ }\href {\doibase
  10.1088/0004-637X/697/1/610} {\bibfield  {journal} {\bibinfo  {journal}
  {\apj}\ }\textbf {\bibinfo {volume} {697}},\ \bibinfo {pages} {610} (\bibinfo
  {year} {2009})},\ \Eprint {http://arxiv.org/abs/0903.2535} {arXiv:0903.2535
  [astro-ph.CO]} \BibitemShut {NoStop}%
\bibitem [{\citenamefont {{Xu}}\ \emph {et~al.}(2009)\citenamefont {{Xu}},
  \citenamefont {{Mao}}, \citenamefont {{Wang}}, \citenamefont {{Springel}},
  \citenamefont {{Gao}}, \citenamefont {{White}}, \citenamefont {{Frenk}},
  \citenamefont {{Jenkins}}, \citenamefont {{Li}},\ and\ \citenamefont
  {{Navarro}}}]{xu2009}%
  \BibitemOpen
  \bibfield  {author} {\bibinfo {author} {\bibfnamefont {D.~D.}\ \bibnamefont
  {{Xu}}}, \bibinfo {author} {\bibfnamefont {S.}~\bibnamefont {{Mao}}},
  \bibinfo {author} {\bibfnamefont {J.}~\bibnamefont {{Wang}}}, \bibinfo
  {author} {\bibfnamefont {V.}~\bibnamefont {{Springel}}}, \bibinfo {author}
  {\bibfnamefont {L.}~\bibnamefont {{Gao}}}, \bibinfo {author} {\bibfnamefont
  {S.~D.~M.}\ \bibnamefont {{White}}}, \bibinfo {author} {\bibfnamefont
  {C.~S.}\ \bibnamefont {{Frenk}}}, \bibinfo {author} {\bibfnamefont
  {A.}~\bibnamefont {{Jenkins}}}, \bibinfo {author} {\bibfnamefont
  {G.}~\bibnamefont {{Li}}}, \ and\ \bibinfo {author} {\bibfnamefont {J.~F.}\
  \bibnamefont {{Navarro}}},\ }\href {\doibase
  10.1111/j.1365-2966.2009.15230.x} {\bibfield  {journal} {\bibinfo  {journal}
  {\mnras}\ }\textbf {\bibinfo {volume} {398}},\ \bibinfo {pages} {1235}
  (\bibinfo {year} {2009})},\ \Eprint {http://arxiv.org/abs/0903.4559}
  {arXiv:0903.4559} \BibitemShut {NoStop}%
\bibitem [{\citenamefont {{Xu}}\ \emph {et~al.}(2010)\citenamefont {{Xu}},
  \citenamefont {{Mao}}, \citenamefont {{Cooper}}, \citenamefont {{Wang}},
  \citenamefont {{Gao}}, \citenamefont {{Frenk}},\ and\ \citenamefont
  {{Springel}}}]{xu2010}%
  \BibitemOpen
  \bibfield  {author} {\bibinfo {author} {\bibfnamefont {D.~D.}\ \bibnamefont
  {{Xu}}}, \bibinfo {author} {\bibfnamefont {S.}~\bibnamefont {{Mao}}},
  \bibinfo {author} {\bibfnamefont {A.~P.}\ \bibnamefont {{Cooper}}}, \bibinfo
  {author} {\bibfnamefont {J.}~\bibnamefont {{Wang}}}, \bibinfo {author}
  {\bibfnamefont {L.}~\bibnamefont {{Gao}}}, \bibinfo {author} {\bibfnamefont
  {C.~S.}\ \bibnamefont {{Frenk}}}, \ and\ \bibinfo {author} {\bibfnamefont
  {V.}~\bibnamefont {{Springel}}},\ }\href {\doibase
  10.1111/j.1365-2966.2010.17235.x} {\bibfield  {journal} {\bibinfo  {journal}
  {\mnras}\ }\textbf {\bibinfo {volume} {408}},\ \bibinfo {pages} {1721}
  (\bibinfo {year} {2010})},\ \Eprint {http://arxiv.org/abs/1004.3094}
  {arXiv:1004.3094} \BibitemShut {NoStop}%
\bibitem [{\citenamefont {{Fadely}}\ and\ \citenamefont
  {{Keeton}}(2012)}]{fadely2012}%
  \BibitemOpen
  \bibfield  {author} {\bibinfo {author} {\bibfnamefont {R.}~\bibnamefont
  {{Fadely}}}\ and\ \bibinfo {author} {\bibfnamefont {C.~R.}\ \bibnamefont
  {{Keeton}}},\ }\href {\doibase 10.1111/j.1365-2966.2011.19729.x} {\bibfield
  {journal} {\bibinfo  {journal} {\mnras}\ }\textbf {\bibinfo {volume} {419}},\
  \bibinfo {pages} {936} (\bibinfo {year} {2012})},\ \Eprint
  {http://arxiv.org/abs/1109.0548} {arXiv:1109.0548} \BibitemShut {NoStop}%
\bibitem [{\citenamefont {{MacLeod}}\ \emph {et~al.}(2013)\citenamefont
  {{MacLeod}}, \citenamefont {{Jones}}, \citenamefont {{Agol}},\ and\
  \citenamefont {{Kochanek}}}]{macleod2013}%
  \BibitemOpen
  \bibfield  {author} {\bibinfo {author} {\bibfnamefont {C.~L.}\ \bibnamefont
  {{MacLeod}}}, \bibinfo {author} {\bibfnamefont {R.}~\bibnamefont {{Jones}}},
  \bibinfo {author} {\bibfnamefont {E.}~\bibnamefont {{Agol}}}, \ and\ \bibinfo
  {author} {\bibfnamefont {C.~S.}\ \bibnamefont {{Kochanek}}},\ }\href
  {\doibase 10.1088/0004-637X/773/1/35} {\bibfield  {journal} {\bibinfo
  {journal} {\apj}\ }\textbf {\bibinfo {volume} {773}},\ \bibinfo {eid} {35}
  (\bibinfo {year} {2013})},\ \Eprint {http://arxiv.org/abs/1212.2166}
  {arXiv:1212.2166 [astro-ph.CO]} \BibitemShut {NoStop}%
\bibitem [{\citenamefont {{Chen}}\ \emph {et~al.}(2003)\citenamefont {{Chen}},
  \citenamefont {{Kravtsov}},\ and\ \citenamefont {{Keeton}}}]{chen2003}%
  \BibitemOpen
  \bibfield  {author} {\bibinfo {author} {\bibfnamefont {J.}~\bibnamefont
  {{Chen}}}, \bibinfo {author} {\bibfnamefont {A.~V.}\ \bibnamefont
  {{Kravtsov}}}, \ and\ \bibinfo {author} {\bibfnamefont {C.~R.}\ \bibnamefont
  {{Keeton}}},\ }\href {\doibase 10.1086/375639} {\bibfield  {journal}
  {\bibinfo  {journal} {\apj}\ }\textbf {\bibinfo {volume} {592}},\ \bibinfo
  {pages} {24} (\bibinfo {year} {2003})},\ \Eprint
  {http://arxiv.org/abs/astro-ph/0302005} {astro-ph/0302005} \BibitemShut
  {NoStop}%
\bibitem [{\citenamefont {{Metcalf}}(2005)}]{metcalf2005a}%
  \BibitemOpen
  \bibfield  {author} {\bibinfo {author} {\bibfnamefont {R.~B.}\ \bibnamefont
  {{Metcalf}}},\ }\href {\doibase 10.1086/431574} {\bibfield  {journal}
  {\bibinfo  {journal} {\apj}\ }\textbf {\bibinfo {volume} {629}},\ \bibinfo
  {pages} {673} (\bibinfo {year} {2005})},\ \Eprint
  {http://arxiv.org/abs/astro-ph/0412538} {astro-ph/0412538} \BibitemShut
  {NoStop}%
\bibitem [{\citenamefont {{Xu}}\ \emph {et~al.}(2012)\citenamefont {{Xu}},
  \citenamefont {{Mao}}, \citenamefont {{Cooper}}, \citenamefont {{Gao}},
  \citenamefont {{Frenk}}, \citenamefont {{Angulo}},\ and\ \citenamefont
  {{Helly}}}]{xu2012}%
  \BibitemOpen
  \bibfield  {author} {\bibinfo {author} {\bibfnamefont {D.~D.}\ \bibnamefont
  {{Xu}}}, \bibinfo {author} {\bibfnamefont {S.}~\bibnamefont {{Mao}}},
  \bibinfo {author} {\bibfnamefont {A.~P.}\ \bibnamefont {{Cooper}}}, \bibinfo
  {author} {\bibfnamefont {L.}~\bibnamefont {{Gao}}}, \bibinfo {author}
  {\bibfnamefont {C.~S.}\ \bibnamefont {{Frenk}}}, \bibinfo {author}
  {\bibfnamefont {R.~E.}\ \bibnamefont {{Angulo}}}, \ and\ \bibinfo {author}
  {\bibfnamefont {J.}~\bibnamefont {{Helly}}},\ }\href {\doibase
  10.1111/j.1365-2966.2012.20484.x} {\bibfield  {journal} {\bibinfo  {journal}
  {\mnras}\ }\textbf {\bibinfo {volume} {421}},\ \bibinfo {pages} {2553}
  (\bibinfo {year} {2012})},\ \Eprint {http://arxiv.org/abs/1110.1185}
  {arXiv:1110.1185} \BibitemShut {NoStop}%
\bibitem [{\citenamefont {{Inoue}}\ and\ \citenamefont
  {{Takahashi}}(2012)}]{inoue-takahashi2012}%
  \BibitemOpen
  \bibfield  {author} {\bibinfo {author} {\bibfnamefont {K.~T.}\ \bibnamefont
  {{Inoue}}}\ and\ \bibinfo {author} {\bibfnamefont {R.}~\bibnamefont
  {{Takahashi}}},\ }\href {\doibase 10.1111/j.1365-2966.2012.21915.x}
  {\bibfield  {journal} {\bibinfo  {journal} {\mnras}\ }\textbf {\bibinfo
  {volume} {426}},\ \bibinfo {pages} {2978} (\bibinfo {year} {2012})},\ \Eprint
  {http://arxiv.org/abs/1207.2139} {arXiv:1207.2139 [astro-ph.CO]} \BibitemShut
  {NoStop}%
\bibitem [{\citenamefont {{Takahashi}}\ and\ \citenamefont
  {{Inoue}}(2014)}]{takahashi-inoue2014}%
  \BibitemOpen
  \bibfield  {author} {\bibinfo {author} {\bibfnamefont {R.}~\bibnamefont
  {{Takahashi}}}\ and\ \bibinfo {author} {\bibfnamefont {K.~T.}\ \bibnamefont
  {{Inoue}}},\ }\href {\doibase 10.1093/mnras/stu328} {\bibfield  {journal}
  {\bibinfo  {journal} {\mnras}\ }\textbf {\bibinfo {volume} {440}},\ \bibinfo
  {pages} {870} (\bibinfo {year} {2014})},\ \Eprint
  {http://arxiv.org/abs/1308.4855} {arXiv:1308.4855 [astro-ph.CO]} \BibitemShut
  {NoStop}%
\bibitem [{\citenamefont {{Inoue}}\ \emph
  {et~al.}(2015{\natexlab{a}})\citenamefont {{Inoue}}, \citenamefont
  {{Takahashi}}, \citenamefont {{Takahashi}},\ and\ \citenamefont
  {{Ishiyama}}}]{inoue-etal2015}%
  \BibitemOpen
  \bibfield  {author} {\bibinfo {author} {\bibfnamefont {K.~T.}\ \bibnamefont
  {{Inoue}}}, \bibinfo {author} {\bibfnamefont {R.}~\bibnamefont
  {{Takahashi}}}, \bibinfo {author} {\bibfnamefont {T.}~\bibnamefont
  {{Takahashi}}}, \ and\ \bibinfo {author} {\bibfnamefont {T.}~\bibnamefont
  {{Ishiyama}}},\ }\href {\doibase 10.1093/mnras/stv194} {\bibfield  {journal}
  {\bibinfo  {journal} {\mnras}\ }\textbf {\bibinfo {volume} {448}},\ \bibinfo
  {pages} {2704} (\bibinfo {year} {2015}{\natexlab{a}})},\ \Eprint
  {http://arxiv.org/abs/1409.1326} {arXiv:1409.1326} \BibitemShut {NoStop}%
\bibitem [{\citenamefont {{Inoue}}\ \emph
  {et~al.}(2015{\natexlab{b}})\citenamefont {{Inoue}}, \citenamefont
  {{Minezaki}}, \citenamefont {{Matsushita}},\ and\ \citenamefont
  {{Chiba}}}]{inoue-minezaki2015}%
  \BibitemOpen
  \bibfield  {author} {\bibinfo {author} {\bibfnamefont {K.~T.}\ \bibnamefont
  {{Inoue}}}, \bibinfo {author} {\bibfnamefont {T.}~\bibnamefont {{Minezaki}}},
  \bibinfo {author} {\bibfnamefont {S.}~\bibnamefont {{Matsushita}}}, \ and\
  \bibinfo {author} {\bibfnamefont {M.}~\bibnamefont {{Chiba}}},\ }\href@noop
  {} {\bibfield  {journal} {\bibinfo  {journal} {ArXiv e-prints}\ } (\bibinfo
  {year} {2015}{\natexlab{b}})},\ \Eprint {http://arxiv.org/abs/1510.00150}
  {arXiv:1510.00150} \BibitemShut {NoStop}%
\bibitem [{\citenamefont {{Xu}}\ \emph {et~al.}(2015)\citenamefont {{Xu}},
  \citenamefont {{Sluse}}, \citenamefont {{Gao}}, \citenamefont {{Wang}},
  \citenamefont {{Frenk}}, \citenamefont {{Mao}}, \citenamefont {{Schneider}},\
  and\ \citenamefont {{Springel}}}]{xu2015}%
  \BibitemOpen
  \bibfield  {author} {\bibinfo {author} {\bibfnamefont {D.}~\bibnamefont
  {{Xu}}}, \bibinfo {author} {\bibfnamefont {D.}~\bibnamefont {{Sluse}}},
  \bibinfo {author} {\bibfnamefont {L.}~\bibnamefont {{Gao}}}, \bibinfo
  {author} {\bibfnamefont {J.}~\bibnamefont {{Wang}}}, \bibinfo {author}
  {\bibfnamefont {C.}~\bibnamefont {{Frenk}}}, \bibinfo {author} {\bibfnamefont
  {S.}~\bibnamefont {{Mao}}}, \bibinfo {author} {\bibfnamefont
  {P.}~\bibnamefont {{Schneider}}}, \ and\ \bibinfo {author} {\bibfnamefont
  {V.}~\bibnamefont {{Springel}}},\ }\href {\doibase 10.1093/mnras/stu2673}
  {\bibfield  {journal} {\bibinfo  {journal} {\mnras}\ }\textbf {\bibinfo
  {volume} {447}},\ \bibinfo {pages} {3189} (\bibinfo {year} {2015})},\ \Eprint
  {http://arxiv.org/abs/1410.3282} {arXiv:1410.3282} \BibitemShut {NoStop}%
\bibitem [{\citenamefont {{Inoue}}(2016)}]{inoue2016}%
  \BibitemOpen
  \bibfield  {author} {\bibinfo {author} {\bibfnamefont {K.~T.}\ \bibnamefont
  {{Inoue}}},\ }\href@noop {} {\bibfield  {journal} {\bibinfo  {journal} {ArXiv
  e-prints}\ } (\bibinfo {year} {2016})},\ \Eprint
  {http://arxiv.org/abs/1601.04414} {arXiv:1601.04414} \BibitemShut {NoStop}%
\bibitem [{\citenamefont {Miranda}\ and\ \citenamefont
  {Maccio}(2007)}]{miranda2007}%
  \BibitemOpen
  \bibfield  {author} {\bibinfo {author} {\bibfnamefont {M.}~\bibnamefont
  {Miranda}}\ and\ \bibinfo {author} {\bibfnamefont {A.~V.}\ \bibnamefont
  {Maccio}},\ }\href@noop {} {\bibfield  {journal} {\bibinfo  {journal}
  {\mnras}\ }\textbf {\bibinfo {volume} {382}},\ \bibinfo {pages} {1225}
  (\bibinfo {year} {2007})},\ \bibinfo {note} {235RS Times Cited:26 Cited
  References Count:58}\BibitemShut {NoStop}%
\bibitem [{\citenamefont {Ade}\ \emph {et~al.}(2014)\citenamefont {Ade} \emph
  {et~al.}}]{Ade:2013zuv}%
  \BibitemOpen
  \bibfield  {author} {\bibinfo {author} {\bibfnamefont {P.~A.~R.}\
  \bibnamefont {Ade}} \emph {et~al.} (\bibinfo {collaboration} {Planck}),\
  }\href {\doibase 10.1051/0004-6361/201321591} {\bibfield  {journal} {\bibinfo
   {journal} {\aap}\ }\textbf {\bibinfo {volume} {571}},\ \bibinfo {pages}
  {A16} (\bibinfo {year} {2014})},\ \Eprint {http://arxiv.org/abs/1303.5076}
  {arXiv:1303.5076 [astro-ph.CO]} \BibitemShut {NoStop}%
\bibitem [{\citenamefont {Lewis}\ \emph {et~al.}(2000)\citenamefont {Lewis},
  \citenamefont {Challinor},\ and\ \citenamefont {Lasenby}}]{Lewis:1999bs}%
  \BibitemOpen
  \bibfield  {author} {\bibinfo {author} {\bibfnamefont {A.}~\bibnamefont
  {Lewis}}, \bibinfo {author} {\bibfnamefont {A.}~\bibnamefont {Challinor}}, \
  and\ \bibinfo {author} {\bibfnamefont {A.}~\bibnamefont {Lasenby}},\ }\href
  {\doibase 10.1086/309179} {\bibfield  {journal} {\bibinfo  {journal} {\apj}\
  }\textbf {\bibinfo {volume} {538}},\ \bibinfo {pages} {473} (\bibinfo {year}
  {2000})},\ \Eprint {http://arxiv.org/abs/astro-ph/9911177}
  {arXiv:astro-ph/9911177 [astro-ph]} \BibitemShut {NoStop}%
\bibitem [{\citenamefont {Kamada}\ \emph {et~al.}(2013)\citenamefont {Kamada},
  \citenamefont {Yoshida}, \citenamefont {Kohri},\ and\ \citenamefont
  {Takahashi}}]{Kamada:2013sh}%
  \BibitemOpen
  \bibfield  {author} {\bibinfo {author} {\bibfnamefont {A.}~\bibnamefont
  {Kamada}}, \bibinfo {author} {\bibfnamefont {N.}~\bibnamefont {Yoshida}},
  \bibinfo {author} {\bibfnamefont {K.}~\bibnamefont {Kohri}}, \ and\ \bibinfo
  {author} {\bibfnamefont {T.}~\bibnamefont {Takahashi}},\ }\href {\doibase
  10.1088/1475-7516/2013/03/008} {\bibfield  {journal} {\bibinfo  {journal}
  {JCAP}\ }\textbf {\bibinfo {volume} {1303}},\ \bibinfo {pages} {008}
  (\bibinfo {year} {2013})},\ \Eprint {http://arxiv.org/abs/1301.2744}
  {arXiv:1301.2744 [astro-ph.CO]} \BibitemShut {NoStop}%
\bibitem [{\citenamefont {Springel}(2005)}]{Springel:2005mi}%
  \BibitemOpen
  \bibfield  {author} {\bibinfo {author} {\bibfnamefont {V.}~\bibnamefont
  {Springel}},\ }\href {\doibase 10.1111/j.1365-2966.2005.09655.x} {\bibfield
  {journal} {\bibinfo  {journal} {\mnras}\ }\textbf {\bibinfo {volume} {364}},\
  \bibinfo {pages} {1105} (\bibinfo {year} {2005})},\ \Eprint
  {http://arxiv.org/abs/astro-ph/0505010} {arXiv:astro-ph/0505010 [astro-ph]}
  \BibitemShut {NoStop}%
\bibitem [{\citenamefont {{Koopmans}}\ \emph {et~al.}(2003)\citenamefont
  {{Koopmans}}, \citenamefont {{Biggs}}, \citenamefont {{Blandford}},
  \citenamefont {{Browne}}, \citenamefont {{Jackson}}, \citenamefont {{Mao}},
  \citenamefont {{Wilkinson}}, \citenamefont {{de Bruyn}},\ and\ \citenamefont
  {{Wambsganss}}}]{koopmans2003}%
  \BibitemOpen
  \bibfield  {author} {\bibinfo {author} {\bibfnamefont {L.~V.~E.}\
  \bibnamefont {{Koopmans}}}, \bibinfo {author} {\bibfnamefont
  {A.}~\bibnamefont {{Biggs}}}, \bibinfo {author} {\bibfnamefont {R.~D.}\
  \bibnamefont {{Blandford}}}, \bibinfo {author} {\bibfnamefont {I.~W.~A.}\
  \bibnamefont {{Browne}}}, \bibinfo {author} {\bibfnamefont {N.~J.}\
  \bibnamefont {{Jackson}}}, \bibinfo {author} {\bibfnamefont {S.}~\bibnamefont
  {{Mao}}}, \bibinfo {author} {\bibfnamefont {P.~N.}\ \bibnamefont
  {{Wilkinson}}}, \bibinfo {author} {\bibfnamefont {A.~G.}\ \bibnamefont {{de
  Bruyn}}}, \ and\ \bibinfo {author} {\bibfnamefont {J.}~\bibnamefont
  {{Wambsganss}}},\ }\href {\doibase 10.1086/377434} {\bibfield  {journal}
  {\bibinfo  {journal} {\apj}\ }\textbf {\bibinfo {volume} {595}},\ \bibinfo
  {pages} {712} (\bibinfo {year} {2003})},\ \Eprint
  {http://arxiv.org/abs/astro-ph/0302189} {astro-ph/0302189} \BibitemShut
  {NoStop}%
\bibitem [{\citenamefont {{Sluse}}\ \emph {et~al.}(2012)\citenamefont
  {{Sluse}}, \citenamefont {{Chantry}}, \citenamefont {{Magain}}, \citenamefont
  {{Courbin}},\ and\ \citenamefont {{Meylan}}}]{sluse2012}%
  \BibitemOpen
  \bibfield  {author} {\bibinfo {author} {\bibfnamefont {D.}~\bibnamefont
  {{Sluse}}}, \bibinfo {author} {\bibfnamefont {V.}~\bibnamefont {{Chantry}}},
  \bibinfo {author} {\bibfnamefont {P.}~\bibnamefont {{Magain}}}, \bibinfo
  {author} {\bibfnamefont {F.}~\bibnamefont {{Courbin}}}, \ and\ \bibinfo
  {author} {\bibfnamefont {G.}~\bibnamefont {{Meylan}}},\ }\href {\doibase
  10.1051/0004-6361/201015844} {\bibfield  {journal} {\bibinfo  {journal}
  {Astronomy and Astrophysics}\ }\textbf {\bibinfo {volume} {538}},\ \bibinfo
  {eid} {A99} (\bibinfo {year} {2012})},\ \Eprint
  {http://arxiv.org/abs/1112.0005} {arXiv:1112.0005 [astro-ph.CO]} \BibitemShut
  {NoStop}%
\bibitem [{\citenamefont {{Biggs}}\ \emph {et~al.}(2004)\citenamefont
  {{Biggs}}, \citenamefont {{Browne}}, \citenamefont {{Jackson}}, \citenamefont
  {{York}}, \citenamefont {{Norbury}}, \citenamefont {{McKean}},\ and\
  \citenamefont {{Phillips}}}]{biggs2004}%
  \BibitemOpen
  \bibfield  {author} {\bibinfo {author} {\bibfnamefont {A.~D.}\ \bibnamefont
  {{Biggs}}}, \bibinfo {author} {\bibfnamefont {I.~W.~A.}\ \bibnamefont
  {{Browne}}}, \bibinfo {author} {\bibfnamefont {N.~J.}\ \bibnamefont
  {{Jackson}}}, \bibinfo {author} {\bibfnamefont {T.}~\bibnamefont {{York}}},
  \bibinfo {author} {\bibfnamefont {M.~A.}\ \bibnamefont {{Norbury}}}, \bibinfo
  {author} {\bibfnamefont {J.~P.}\ \bibnamefont {{McKean}}}, \ and\ \bibinfo
  {author} {\bibfnamefont {P.~M.}\ \bibnamefont {{Phillips}}},\ }\href
  {\doibase 10.1111/j.1365-2966.2004.07701.x} {\bibfield  {journal} {\bibinfo
  {journal} {\mnras}\ }\textbf {\bibinfo {volume} {350}},\ \bibinfo {pages}
  {949} (\bibinfo {year} {2004})},\ \Eprint
  {http://arxiv.org/abs/astro-ph/0402128} {astro-ph/0402128} \BibitemShut
  {NoStop}%
\bibitem [{\citenamefont {{Lagattuta}}\ \emph {et~al.}(2010)\citenamefont
  {{Lagattuta}}, \citenamefont {{Auger}},\ and\ \citenamefont
  {{Fassnacht}}}]{lagattuta2010}%
  \BibitemOpen
  \bibfield  {author} {\bibinfo {author} {\bibfnamefont {D.~J.}\ \bibnamefont
  {{Lagattuta}}}, \bibinfo {author} {\bibfnamefont {M.~W.}\ \bibnamefont
  {{Auger}}}, \ and\ \bibinfo {author} {\bibfnamefont {C.~D.}\ \bibnamefont
  {{Fassnacht}}},\ }\href {\doibase 10.1088/2041-8205/716/2/L185} {\bibfield
  {journal} {\bibinfo  {journal} {\apj}\ }\textbf {\bibinfo {volume} {716}},\
  \bibinfo {pages} {L185} (\bibinfo {year} {2010})},\ \Eprint
  {http://arxiv.org/abs/0912.2344} {arXiv:0912.2344 [astro-ph.CO]} \BibitemShut
  {NoStop}%
\bibitem [{\citenamefont {{Kormann}}\ \emph {et~al.}(1994)\citenamefont
  {{Kormann}}, \citenamefont {{Schneider}},\ and\ \citenamefont
  {{Bartelmann}}}]{kormann1994}%
  \BibitemOpen
  \bibfield  {author} {\bibinfo {author} {\bibfnamefont {R.}~\bibnamefont
  {{Kormann}}}, \bibinfo {author} {\bibfnamefont {P.}~\bibnamefont
  {{Schneider}}}, \ and\ \bibinfo {author} {\bibfnamefont {M.}~\bibnamefont
  {{Bartelmann}}},\ }\href@noop {} {\bibfield  {journal} {\bibinfo  {journal}
  {Astronomy and Astrophysics}\ }\textbf {\bibinfo {volume} {284}},\ \bibinfo
  {pages} {285} (\bibinfo {year} {1994})}\BibitemShut {NoStop}%
\end{thebibliography}%

\end{document}